\documentclass[usenatbib,letterpaper]{mn2e}
\input{psfig}
\usepackage{natbib} 
\usepackage{varioref}

\begin{document} 
\def\lsim{\mathrel{\hbox{\rlap{\hbox{\lower4pt\hbox{$\sim$}}}\hbox{$<$}}}}
\def\gsim{\mathrel{\hbox{\rlap{\hbox{\lower4pt\hbox{$\sim$}}}\hbox{$>$}}}}
\def\simlt{\mathrel{\rlap{\lower 3pt\hbox{$\sim$}}
        \raise 2.0pt\hbox{$<$}}}
\def\simgt{\mathrel{\rlap{\lower 3pt\hbox{$\sim$}}
        \raise 2.0pt\hbox{$>$}}}

\title[AGN In Cosmological Simulations]
{Active Galactic Nuclei In Cosmological Simulations - I. Formation of black holes and spheroids through mergers}
\author[A. Cattaneo, J. Blaizot, J. Devriendt, B. Guiderdoni]
{A.~Cattaneo $^{1,2,3}$, J.~Blaizot $^4$, J.~Devriendt $^5$, B.~Guiderdoni $^{1,5}$\\
\\
$^1$Institut d'Astrophysique de Paris, 98bis Boulevard Arago, 75014 Paris, France\\
$^2$Racah Institute of Physics, The Hebrew University, 91904 Jerusalem, Israel\\
$^3$Astrophysikalisches Institut Potsdam, an der Sternwarte 16, 14482 Potsdam, Germany\\
$^4$Max-Planck-Institut f\"ur Astrophysik, Karl-Schwarzschild-Str. 1, 85740 Garching bei M\"unchen, Germany\\
$^5$Centre de Recherche Astronomique de Lyon, 9 Avenue Charles Andr\'e, 69561 St-Genis-Laval Cedex, France\\}

\maketitle 
\begin{abstract}

This is the first paper of a series on the methods and results of the
Active Galactic Nuclei In Cosmological Simulations (AGNICS) project,
which incorporates the physics of AGN into GalICS,
a galaxy formation model that combines
large cosmological N-body simulations of dark matter hierarchical clustering and
a semi-analytic approach to the physics of the baryons.
The project explores the quasar-galaxy link in a cosmological perspective, 
in response to growing observational evidence for a close relation between
supermassive black holes (SMBHs) and spheroids.
The key problems are the quasar fuelling mechanism, the origin of the BH to bulge mass relation,
the causal and chronological link between BH growth and galaxy formation,
the properties of quasar hosts and the role of AGN feedback in galaxy formation. 

This first paper has two goals. The first is 
to describe the general structure and assumptions that provide the framework for the AGNICS series. The second
is to apply AGNICS to studying the joint formation of SMBHs and spheroids in galaxy mergers. 
We investigate under what conditions this scenario can reproduce the
local distribution of SMBHs in nearby galaxies and the evolution of the quasar population.

AGNICS contains two star formation modes: 
a quiescent mode in discs and a starburst mode in proto-spheroids, the latter triggered by mergers and disc instabilities.
Here we assume that BH growth is linked to the starburst mode.
The simplest version of this scenario, in which the black hole accretion rate 
$\dot{M}_\bullet$ and the star formation rate in the starburst component $\dot{M}_{\rm *burst}$
are simply related by a constant of proportionality,
does not to reproduce the cosmic evolution of the quasar population.
A model in which $\dot{M}_\bullet\propto\rho_{\rm burst}^\zeta\dot{M}_{\rm *burst}$,
where $\rho_{\rm burst}$ is the density of the gas in the starburst and $\zeta\simeq 0.5$,
can explain the evolution of the quasar luminosity function in B-band and X-rays
(taking into account the presence of obscured AGN inferred from X-ray studies).
The scatter and the tilt that this model introduces in the BH-to-bulge mass relation are within the observational
constraints. The model predicts that the quasar contribution increases with the total bolometric luminosity and that,
for a given bulge mass, the most massive black holes are in the bulges with the oldest stars.

\end{abstract}

\newpage
\section{Introduction}
\subsection{The connection between quasars and galaxies: observational evidence and open problems}

Several observational and theoretical arguments suggest a direct link between quasars and galaxy formation.

In `Galactic Nuclei as Collapsed Old Quasars', \citet{lynden-bell69} proposed that a quasar phase is part of normal galaxy evolution
and predicted the presence of supermassive black holes in galaxy cores.
The detection of massive dark objects in the nuclei of nearby galaxies has transformed 
this speculation into an observational fact.
The relation between the black hole mass and the mass, luminosity and velocity dispersion of the host spheroid
\citep{magorrian_etal98,ferrarese_merritt00,gebhardt_etal00,merritt_ferrarese01,tremaine_etal02,marconi_hunt03,haering_rix04}
together with the chronological agreement between the quasar epoch ($z\sim 2-3$) and the stellar
ages of early-type galaxies \citep{cattaneo_bernardi03}
prove that black hole growth mechanisms are directly related to the origin of the bulge component.

This relation opens two kinds of questions.
The first one concerns the triggering mechanism that activates black hole growth (what starts the accretion).
The second concerns the mechanism that determines the final black hole mass 
(e.g. what terminates the accretion).

The first attempts to explain the link between quasars and galaxy formation were in the context of
the monolithic collapse model \citep{eggen_etal62}.
Matter with low angular momentum falls to the centre first forming the galactic 
nucleus and the bulge, while matter with high angular momentum slowly settles into a disc.

\citet{toomre_toomre72} were the pioneers of computer simulations of galaxy interactions.
They argued that galaxy mergers can drive a sudden supply of gas 
into the nuclear region while producing a morphological transformation of spiral galaxies into ellipticals.
Smoothed-particle-hydrodynamics simulations of galaxy mergers confirmed their intuition
\citep{barnes_hernquist91,barnes_hernquist96,mihos_hernquist94,mihos_hernquist96,springel00}.
The impressive results of hydrodynamic simulations together with the
widespread and successful use of the merger model in 
semi-analytic models of hierarchical galaxy formation (see below) contributed to a paradigm shift from 
the monolithic to the merger scenario. 

From the point of view of AGN fuelling, the fundamental points of the merging scenario are that
AGN are fuelled with cold gas from the discs of the merging galaxies and that 
black hole growth is conditional to a triggering process.
A process with a trigger and a short intrinsic duration is consistent with the episodic nature of AGN
(in the monolithic collapse model, the brief life time of quasars was attributed to the rapid consumption of low 
angular momentum gas).

In relation to the mechanism that determines the black hole mass, two scenarios are possible.
The first is that black hole growth is determined by fuel availability.
In this case, the quasar switches off when stars have consumed all the gas (e.g. \citealp{kauffmann_haehnelt00}).
In the second scenario, black hole growth is self-regulated through mechanical 
(jets or winds, \citealp{omma_etal04} and references therein) or 
radiative (Compton heating, \citealp{ciotti_ostriker97,ciotti_ostriker01}) feedback.
In this second case, supermassive black holes may be an essential ingredient of
galaxy formation and not just a mere by-product \citep{silk_rees98,granato_etal04,dimatteo_etal05}.

\subsection{Cosmological models of the formation of quasars and galaxies}

The fuelling mechanism, the timing of the active phase in relation to the host
galaxy evolution and the mechanism that limits the accretion are the three main open problems
in relation to the link between quasars and galaxies.

There are two ways of approaching these problems. 
The first is to observe and model in great detail a small sample of individual
objects. The second is to test if specific assumptions can reproduce the population
properties of quasars and galaxies in a cosmological volume.
Semi-analytic models are a powerful tool for this second type of study.

\citet{white_rees78} and \citet{efstathiou_rees88} were the pioneer of galaxy formation and quasar 
formation in a cosmological scenario, in which structures
form through gravitational instability and hierarchical clustering of primordial density fluctuation.
Semi-analytic models of galaxy formation have given a substantial contribution towards a more detailed
comprehension of galaxy formation in a cosmological scenario. The field is 
now so vast that it is no longer possible to cite all authors who contributed to its development. 
The most advanced models are those of the research groups in Durham (e.g. \citealp{cole_etal00} and 
references therein), Munich (\citealp{kauffmann_etal99} and references therein; 
\citealp{kauffmann_haehnelt00}), Paris \citep{hatton_etal03}
and Santa Cruz \citep{somerville_primack99,somerville_etal01}.
Conceptually, a semi-analytic model is structured into two steps. In the first one, Press-Schechter theory
\citep{press_schechter74} or N-body simulations (i.e \citealp{kauffmann_etal99})
are used to follow gravitational clustering in the dark
matter component. The aim of this first step is to construct merger trees for a sample of dark matter haloes,
which are thought to represent the Universe. The second step is to use  semi-analytic prescriptions  
to follow the physics of the baryons (cooling, star formation, stellar evolution, feedback, chemical 
enrichment, reprocessing of light by dust, mergers, bar instabilities, etc.) in the merger trees.
Models in which merger trees are built by using N-body simulations, such as the GalICS 
(Galaxies In Cosmological Simulations) model by \citet{hatton_etal03}, are sometimes called hybrid models
to distinguish them from pure semi-analytic models, in which merger trees are constructed from 
Press-Schechter theory. 

Quasar modellers have followed a similar path
\citep{carlberg90,haehnelt_rees93,katz_etal94,maehoenen_etal95,bi_fang97,haiman_loeb98,cattaneo_etal99,
nitta99,menou_haiman99,percival_miller99,haiman_menou00,monaco_etal00,valageas_schaeffer00,cattaneo01,
cattaneo02, haiman_hui01,hatz_etal01,martini_weinberg01,nath_etal02,volonteri_etal02,volonteri_etal03a,hatz_etal03}. 
The assumptions in these papers are different, but most of them follow the common strategy of
modelling the quasar luminosity function from the \citet{press_schechter74} mass function of 
dark matter haloes in combination with various assumptions for the probability that a halo of mass 
$M_{\rm halo}$ at redshift $z$ contains a quasar with luminosity $L_{\rm QSO}$.
These studies do not treat the complicated physics of galaxy formation and AGN fuelling, but
show the fundamental scaling relations that a physical model has to satisfy.

The next logical development is the merging of the two research programmes
achieved by including quasars into semi-analytic models of galaxy formation
\citep{kauffmann_haehnelt00,kauffmann_haehnelt02,haehnelt_kauffmann00,
haehnelt_kauffmann02,enoki_etal03,dimatteo_etal03}.
From an AGN perspective, this development allows a more physical modelling and thus a more physical 
understanding, while it reduces the number of arbitrary assumptions that the modeller can make. 
E.g. merging rates cannot be changed without changing galaxy morphologies, colours, luminosity function, etc. 
Of course, that means that more hypotheses and parameters go into the model,
but the current data allow to constrain them.
From a galaxy formation perspective, AGN may provide an answer to two long
standing problems:
the overcooling problem in massive haloes (cD galaxies are too many, too massive and too blue; e.g.
\citealp{hatton_etal03} and references therein)
and the difficulty of reproducing sub-millimetre counts without ending up with galaxies that are too bright 
at the present cosmic epoch (e.g. \citealp{devriendt_guiderdoni00}).

\subsection{The AGNICS project}

The Active Galactic Nuclei In Cosmological Simulations (AGNICS) project 
started as an extension of the GalICS (Galaxies In Cosmological Simulations) hybrid 
galaxy formation model \citep{hatton_etal03}.
AGNICS was developed to investigate some aspects of the link between quasars and galaxy formation
in a broad and coherent context.
Combined models of quasars and galaxy formation are in a qualitative agreement
with the cosmic evolution of the quasar population
\citep{kauffmann_haehnelt00}, but do not reproduce the quantitative decrease of the number
of bright quasars at low redshift.
Secondly, most of the AGN population is optically unseen
\citep{ueda_etal03,sazonov_etal04}, but the impact of obscured AGN on 
models of the evolution of the quasar luminosity function
has not been explored yet.
Moreover, an interesting by-product of the AGNICS project is the possibility to use AGNICS as a tool 
for generating a virtual sky of quasars and galaxies, which may assist the planning of forthcoming observations.
The virtual sky may be used to train automatic recognition algorithms 
and test sources of bias or incompleteness in observational studies.

The structure of this first paper is as follows.
In Section 2 we recall the main assumptions and features of the
GalICS galaxy formation model used to develop the AGNICS model 
while we refer to \citet{hatton_etal03} for a more detailed presentation.
We give particular importance to the origin of galaxy morphologies because a correct modelling of spheroids
is essential to derive meaningful results for black holes and AGN.
In Section 3 we describe the method that AGNICS uses to model black hole growth and AGN.
Thereafter, we distinguish between two main models: 
the simplest possible model, in which the black hole accretion rate and the starburst rate
are simply related by a constant of proportionality (we call this the basic model), and the reference model
(the model which allows us to find the best agreement with the data).
In Section 4 we show that the basic model does not reproduce the low redshift decrease of
the comoving density of bright quasars, but we also show that we can modify the basic model 
by changing the expression for the black hole accretion rate and introducing
a factor proportional to a power of starburst density
($\dot{M}_\bullet\propto\rho_{\rm burst}^\zeta\dot{M}_{\rm *burst}$,
where $\dot{M}_\bullet$ is the black hole accretion rate, $\dot{M}_{\rm *burst}$ is the star formation rate in the 
starburst component and $\rho_{\rm burst}$ is the density of the gas in the starburst).
For $\zeta\simeq 0.5$, the modified model (the reference model) satisfies the observational constraints deriving from optical 
\citep{wolf_etal03,croom_etal04} and X-ray \citep{ueda_etal03} observations of AGN and from the scatter in the 
black hole-to-bulge mass relation \citep{marconi_hunt03,haering_rix04}.
In the Conclusion (Section 5), we discuss what we can learn from the work presented in this article and
which questions that are still open.
The study of quasar hosts is postponed to a future publication of the AGNICS series.

\section{The GalICS galaxy formation model}

GalICS is a model of hierarchical galaxy formation which combines high resolution cosmological simulations to
describe the dark matter content of the Universe with semi-analytic prescriptions to follow the physics of
the baryonic component.
The modules that enter the GalICS model are the cosmology (to generate the merger trees),
the cooling model (to follow cooling of hot gas in dark matter haloes),
the model for the galaxy internal structure and dynamics,
the merger model (merging rates and effect of mergers on galactic morphologies),
the star formation and stellar evolution model (including feedback and metal enrichment) and 
the spectral evolution model (stellar spectra, extinction and dust thermal emission). 
We begin this Section by describing the N-body simulation used to generate the merger trees.
Then we progress into explaining the main physical assumptions that enter each functional unit.

\subsection{The dark matter simulation}

\subsubsection{Simulation parameters} 
The cosmological N-body simulation used to construct the merger trees
was carried out by using the parallel tree-code developed by Ninin (1999).
This simulation was run for a flat cold dark matter model with a cosmological constant of
$\Omega_{\Lambda}=0.667$.
The simulated volume is a cube of side $L_{\rm box}=100h_{100}^{-1}$Mpc, with 
$h_{100}\equiv H_0/100{\rm\,km\,s}^{-1}=0.667$, 
containing $256^3$ particles of mass $8.3\times 10^9$M$_{\odot}$ ($H_0$ is the Hubble constant).
The smoothing length is of 29.29 kpc. The cold dark matter power spectrum was normalised in
agreement with the present day abundance of rich clusters ($\sigma_8 =0.88$).
The simulation produced 100 snapshots spaced logarithmically in the expansion factor 
$(1+z)^{-1}$ from $z=35.59$ to $z=0$.

\subsubsection{Halo identification}
On each snapshot a friend-of-friend algorithm was run to identify
virialised groups of more than 20 particles. The minimum mass of a
dark matter halo is therefore $1.66\times 10^{11}\,M_{\odot}$. 
For each halo we compute a set of properties, which include the position and the velocity of the centre of
mass, the kinetic and the potential energy, the spin parameter and the tensor of inertia.

We fit a triaxial ellipsoid to each dark matter halo identified in the snapshot. 
The semi-axes of the ellipsoid are in a ratio determined by the ratio between the three different components of the inertia tensor. 
We shrink the ellipsoid until the inner region satisfies the virial theorem.
The matter within this virial  volume determines the virial mass.
The virial radius is the radius of a sphere whose volume is equal to the virial volume and
the virial density is equal to the virial mass divided by the virial volume.

When passing from N-body to semi-analytic simulations, we idealise dark matter haloes as
singular isothermal spheres truncated at the virial radius.
The information about each individual halo passed to the semi-analytic model is therefore contained in 
three parameters. They are the virial mass, the virial density (the mean density within the virial radius)
and the spin parameter.

\subsubsection{Merger trees}
The merger trees are computed by linking the haloes identified in each snapshot with their progenitors at the
previous time-step. All predecessors from which a halo has inherited one or more particles are counted as 
progenitors, but only one is the main progenitor, the one that has given the largest particle contribution.
In the same way one can identify a main descendant for each dark matter halo.
The merging histories that we obtain are thus far more complex than those constructed from Press-Schechter
theory as they include the processes of evaporation and fragmentation of dark matter haloes. 
In such situations we assume that all baryons remain in the most massive remnant.

\subsection{From haloes to discs} 

\subsubsection{Gas cooling and infall}
Newly identified haloes are given a mass of hot gas by 
using a universal baryonic fraction of $\Omega_{\rm b}/\Omega_0 = 0.135$. Hot gas is assumed to be shock
heated to the virial temperature of the halo and in hydrostatic
equilibrium with the dark matter potential. 
The density profile of the hot gas is that of a singular isothermal sphere truncated at the virial radius.
The cooling time is calculated from the hot gas density distribution with the \citet{sutherland_dopita93}
cooling function.
The gas that can cool in a time $\Delta t$ is the gas whose cooling time and free-fall time are both lower 
than $\Delta t$. GalICS assumes that gas cooling is accompanied by a simultaneous gas inflow in order to 
maintain the same hot gas density profile at all times.  
The disc radius is $r_{\rm d}=\lambda r_{\rm vir}/\sqrt2$, where $r_{\rm vir}$ is the virial radius and
$\lambda$ is the angular momentum parameter of the halo.
Cooling is inhibited in haloes with $\lambda>0.5$.

We also prevent cooling in haloes that contain a total mass of bulge stars $>10^{11}\,M_\odot$ not to exceed
the observational constraints on the number of massive galaxies.
The theoretical argument behind this solution is that the bulge mass is proportional to the black hole mass
and therefore to the total energy budget of AGN over the life of the halo.
We assume that AGN deposit energy in the intergalactic medium mechanically (through jets or winds) and
radiatively (through hard X-ray photons) and that this input is proportional to the black hole accretion rate.
In reality, the long term outcome of the interaction between the black hole and the intergalactic medium is poorly understood.
In particular, there is no strong physical reason why feedback should become important above a critical black hole mass
(but see \citealp{cattaneo02,dunlop_etal03}).
The cut-off at a bulge mass of $>10^{11}\,M_\odot$ introduced by \citet{hatton_etal03}
and used for the work presented in this article is a temporary measure, whose justification is in its capacity of reproducing the data. 
In fact, all groups working on semi-analytic models of galaxy formation were forced to make arbitrary assumptions to fit the bright
part of the galaxy luminosity function.
Meanwhile, we are looking for a motivated self-consistent approach, which will appear in another paper of the AGNICS series.

\subsubsection{Star formation and stellar evolution}
The star formation rate in the disc is 
\begin{equation}
\label{sfr}
\dot{M}_*={M_{\rm cold}\over\beta_*t_{\rm dyn}}.
\end{equation}
Here $M_{\rm cold}$ is the mass of the gas in the disc (all the gas in the disc is cold and all the gas in the halo
is hot) and $t_{\rm dyn}$ is the dynamical time (the time to complete a half rotation at the disc half mass radius).
The parameter $\beta_*$, which determines the efficiency of star formation has a fiducial value of $\beta_*=50$
\citep{guiderdoni_etal98}.
The mass of newly formed stars is distributed according to the \citet{kennicutt83} initial mass function.
Stars are evolved between time-steps using a sub-stepping of at most 1\,Myr. During each sub-step,
stars release mass and energy into the interstellar medium. Most of the mass comes from the red giant and 
the asymptotic giant branch of stellar evolution, while most of the energy comes from shocks due to
supernova explosions. In GalICS, the enriched material released in the late stages of stellar evolution 
is mixed to the cold phase, while the energy released from supernovae is used to reheat the cold gas and return it to
the hot phase in halo. The reheated gas can also be ejected from the halo if the potential is shallow enough.
The rate of mass loss in the supernova-driven wind that flows out of the disc is directly proportional
to the supernova rate. The details of the model and their justification are found in \citet{hatton_etal03}.

\subsection{Galaxy morphologies}

\subsubsection{Galaxy internal structure}
GalICS models the baryonic part of a galaxy as the sum of three components: the disc, the bulge and the starburst.
These three components are not always present at the same time. 
The fundamental assumption is that all galaxies are born as discs at the centre of a dark matter halo. 
The transformation of disc stars into bulge stars and of disc gas into starbursting gas is is due to 
bar instabilities and mergers. Gas is never added to bulges directly and the only gas in bulges is that coming from stellar mass loss.
The starbursting gas forms a young stellar population that becomes part of the bulge stellar
population when the stars have reached an age of 100$\,$Myr. We do not readjust the bulge radius when this happens.
The disc has an exponential profile, while the bulge and the starburst are described by a \citet{hernquist90} density distribution.
The starburst scale is $r_{\rm burst}=\kappa r_{\rm bulge}$ with $\kappa=0.1$.
This model cannot take into account the presence of more complex morphologies, which are particularly relevant to the case of
starburst galaxies, but has a justification that comes from smoothed particle hydrodynamics simulations of galaxy mergers.
The star formation law (Eq.~\ref{sfr}) has the same form and 
uses the same efficiency parameter $\beta_*$
for all three components when we redefine $M_{\rm cold}$ as the mass of the gas in
the component and $t_{\rm dyn}$ as the dynamical time of the component. For the components described by a Hernquist profile, 
the dynamical time is $t_{\rm dyn}=r_{0.5}/\sigma$, where $r_{0.5}$ is the half mass radius and $\sigma$ is the velocity dispersion
at the half mass radius.

\subsubsection{Disc instabilities}
The formation of a bulge from a disc instability is modelled as follows. A disc is globally stable if $v_{\rm d}<0.7v_{\rm tot}$
(e.g. \citealp{vandenbosch98}).
Here $v_{\rm tot}$ is the circular velocity at the disc half mass radius, computed from the total gravitational potential of the disc,
the bulge, the starburst and the halo,
while $v_{\rm d}$ is the circular velocity computed from the gravitational potential of the disc only.
The bar instability transfers gas from the disc to the starburst and stars from the disc to the bulge until the stability criterion
is fulfilled. 
\begin{figure*}
\noindent
\begin{minipage}{8.6cm}
  \centerline{\hbox{
      \psfig{figure=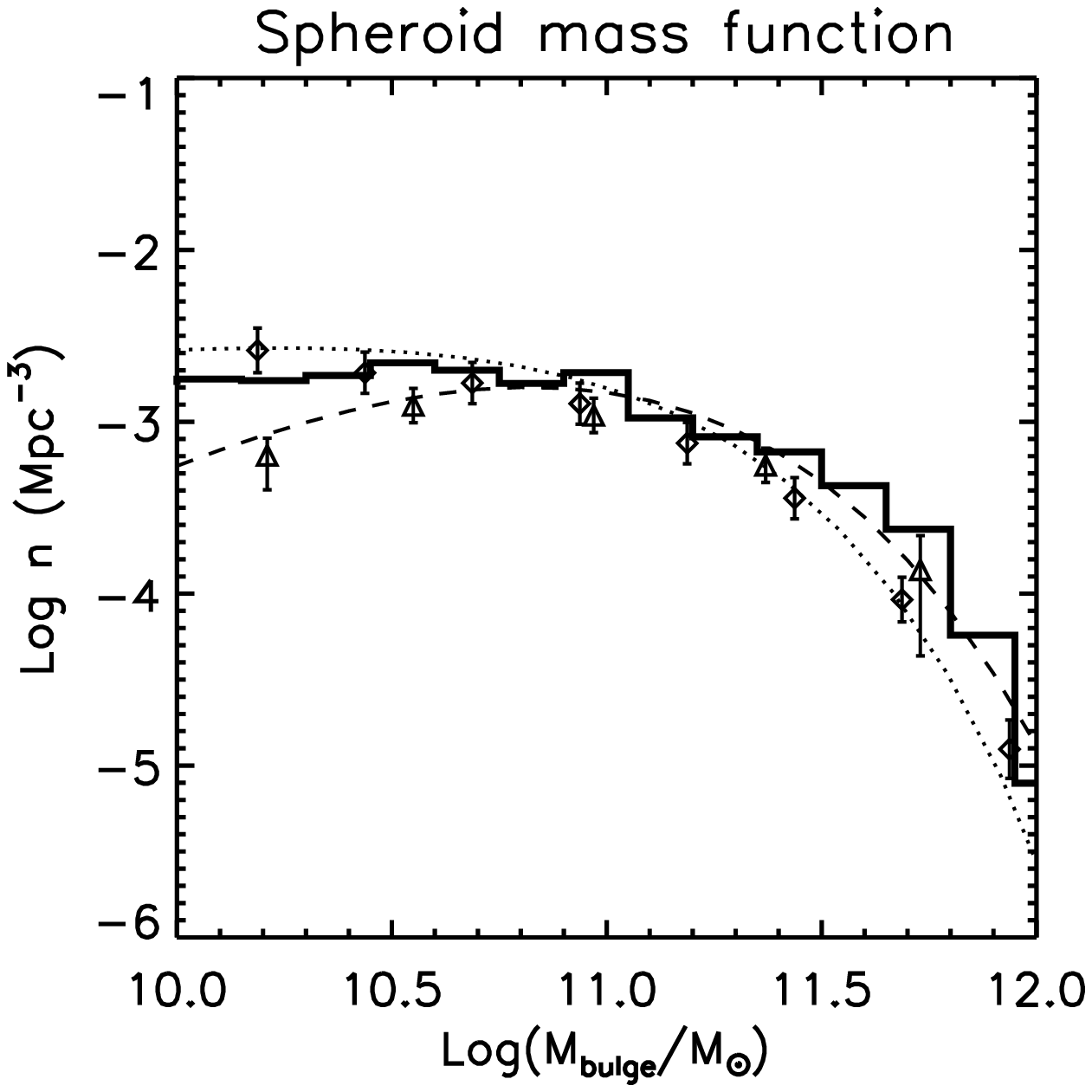,height=8.6cm,angle=0}
  }}
\end{minipage}\    \
\begin{minipage}{8.6cm}
  \centerline{\hbox{
      \psfig{figure=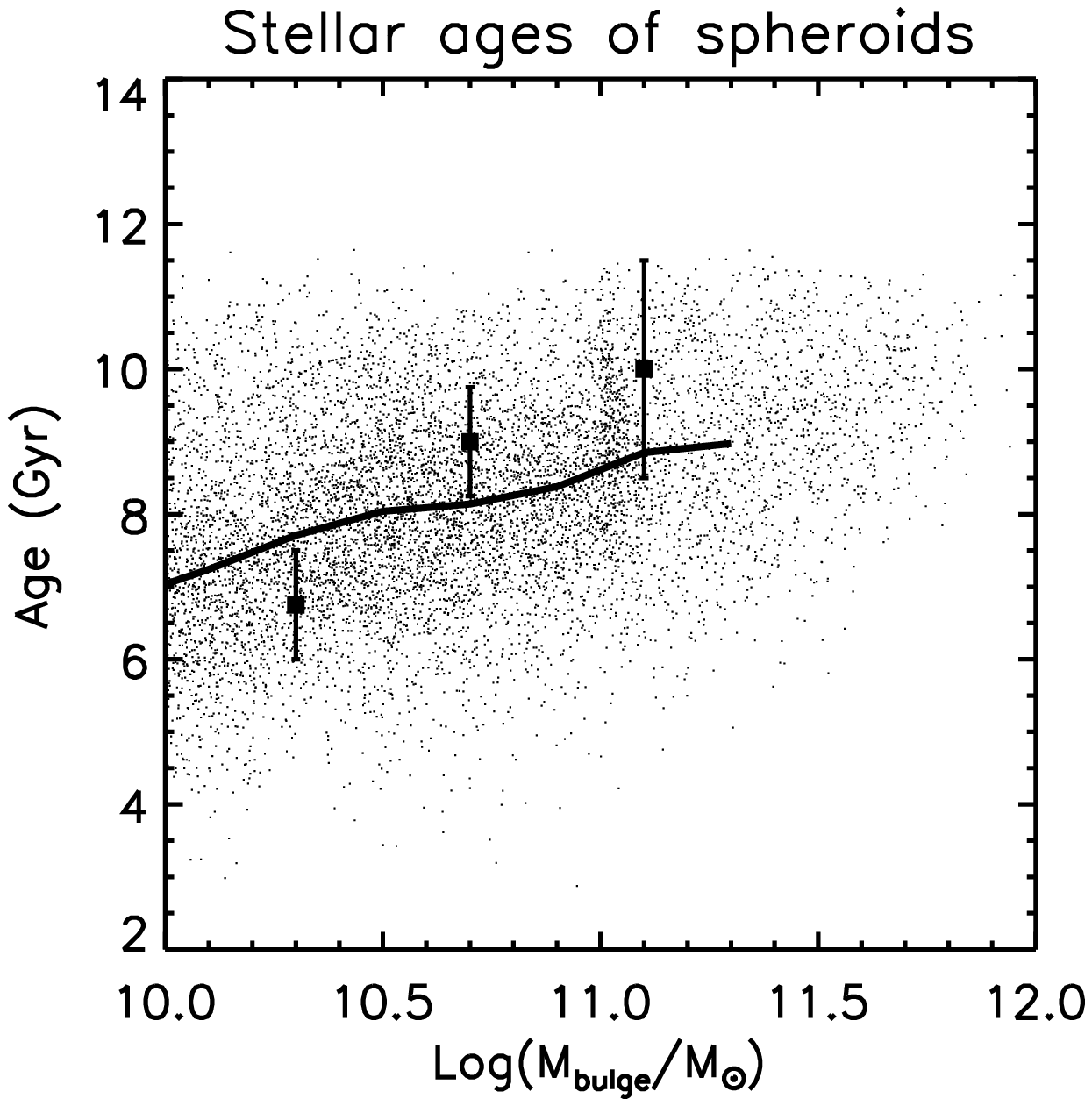,height=8.6cm,angle=0}
  }}
\end{minipage}\    \
\caption{The properties of spheroids in the GalICS model and in the data. The left panel compares the spheroid mass function
predicted by GalICS (solid line) with those inferred from
the 2Mass $K$ luminosity function of  \citet{kochanek_etal01} (diamonds with error bars), 
the  Sloan Digital Sky Survey $r^*$ band luminosity function of \citet{nakamura_etal02} (dotted line),
the $I$ band luminosity function of \citet{benson_etal02} (triangles with error bars) and the
Sloan Digital Sky Survey velocity dispersion distribution in \citet{cattaneo_bernardi03} (dashed line).
The right panel shows the mass - stellar age relation in GalICS (small points; the solid line shows the mean value)
and in the Sloan Digital Sky Survey (\citealp{cattaneo_bernardi03}, points with error bars).}
\end{figure*}
\begin{figure*}
\noindent
\begin{minipage}{8.6cm}
  \centerline{\hbox{
      \psfig{figure=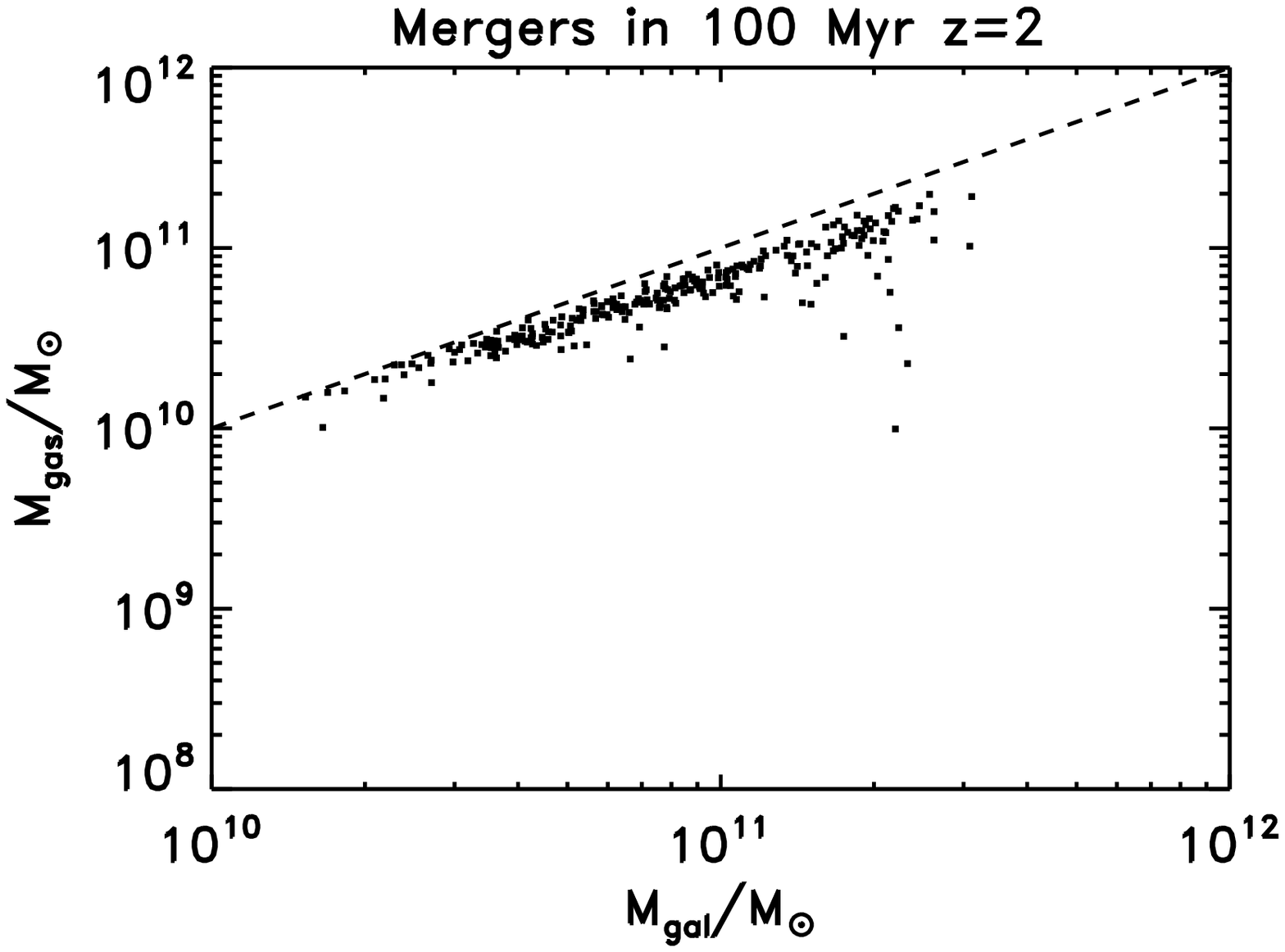,height=6.cm,angle=0}
  }}
\end{minipage}\    \
\begin{minipage}{8.6cm}
  \centerline{\hbox{
      \psfig{figure=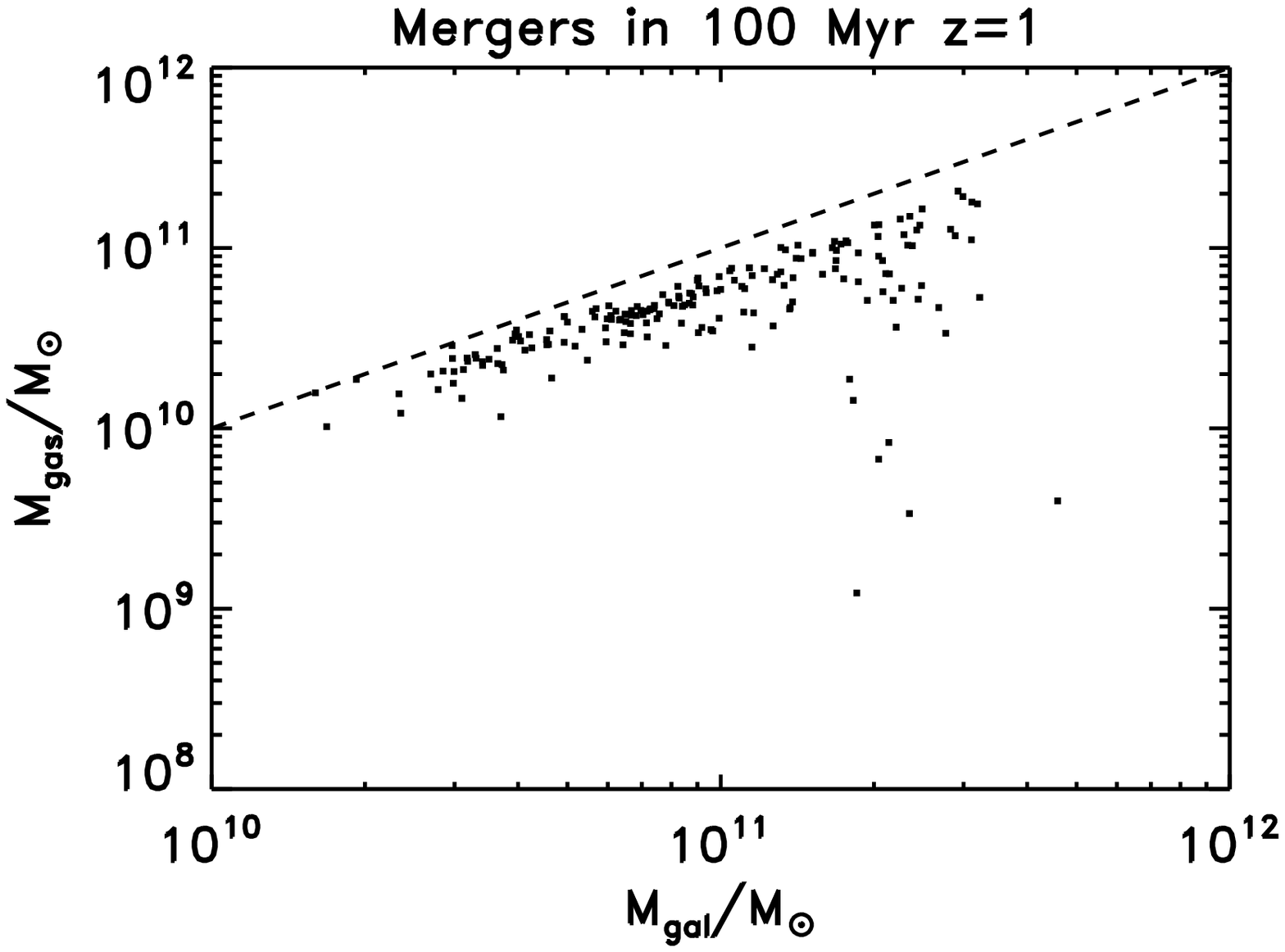,height=6.cm,angle=0}
  }}
\end{minipage}\    \
\begin{minipage}{8.6cm}
  \centerline{\hbox{
      \psfig{figure=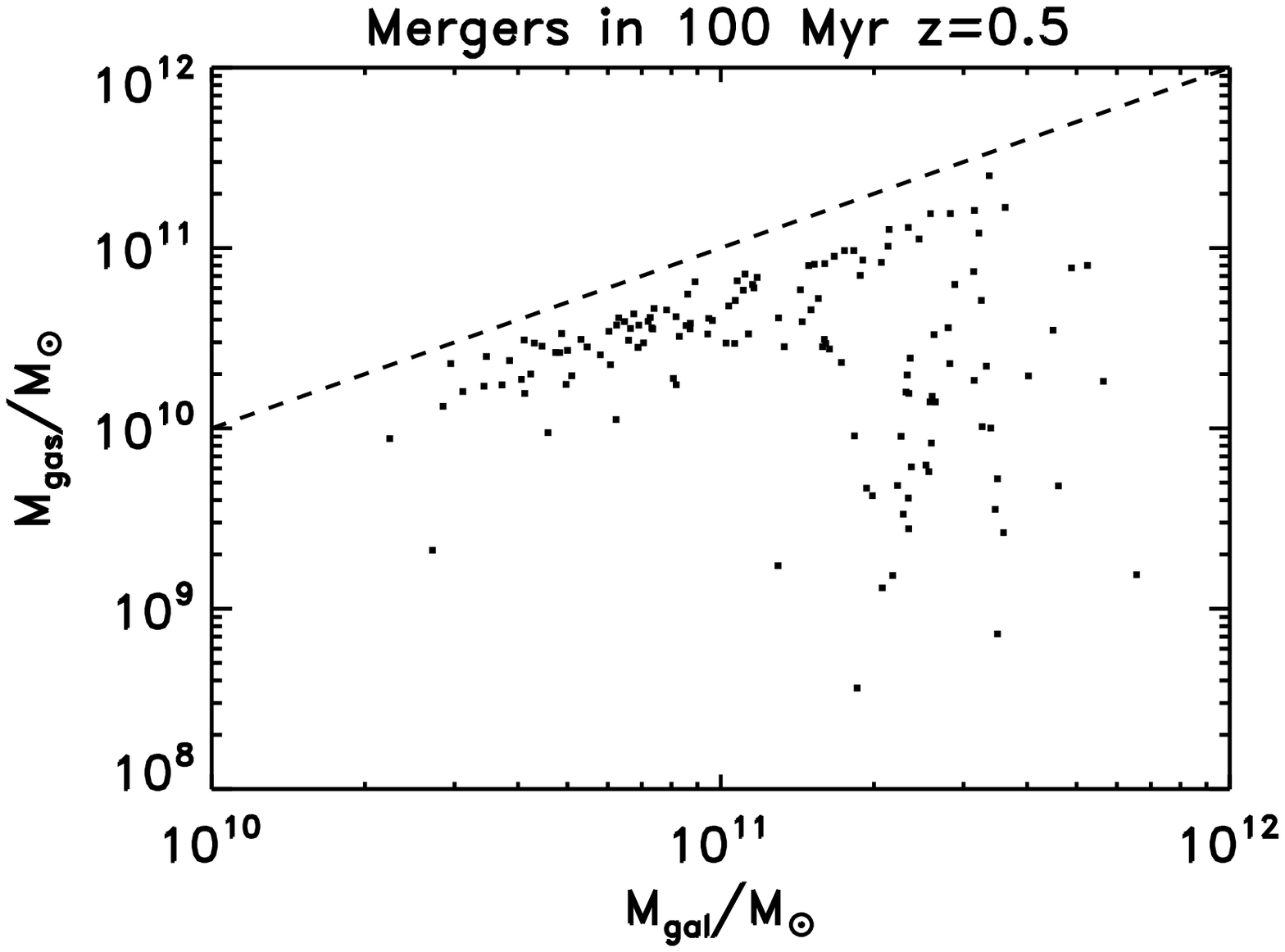,height=6.cm,angle=0}
  }}
\end{minipage}\    \
\begin{minipage}{8.6cm}
  \centerline{\hbox{
      \psfig{figure=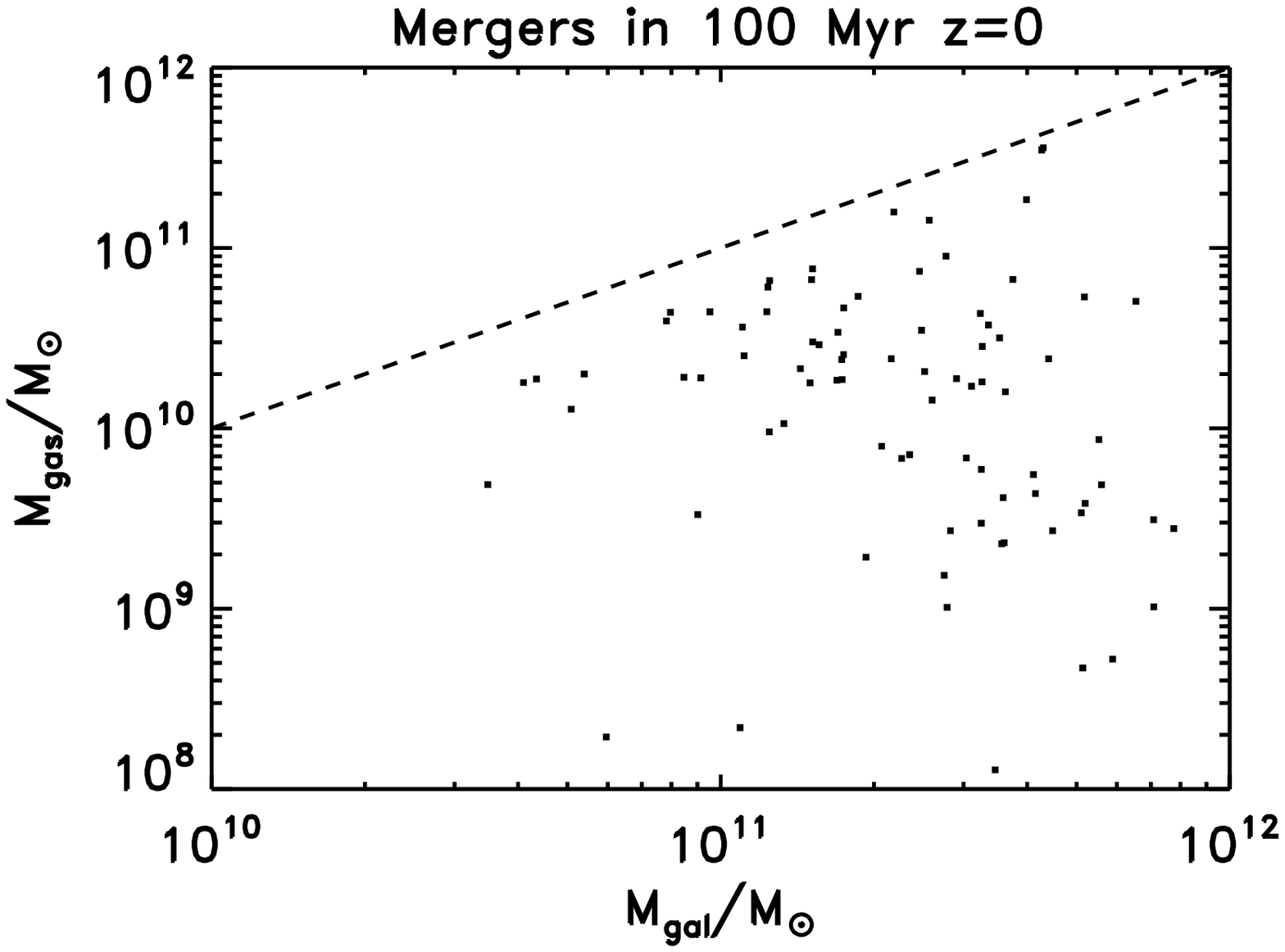,height=6.cm,angle=0}
  }}
\end{minipage}\    \
\caption{Total masses (stars and cold gas, $x$-axis) and gas masses only ($y$-axis)
for mergers that have occurred in four time intervals of 100\,Myr each at redshifts of 2, 1, 0.5 and 0.
The dashed lines correspond to the position on the plots of mergers in which the galaxies are $100\%$ gas.
Notice that the plotting range on the vertical axis is different in each plot. }
\end{figure*}
\subsubsection{Merging rate}
Mergers are the second path to form spheroids. The model for galaxy mergers is articulated into two parts:
the computation of the merging rate and the model for predicting the outcome of mergers when they occur.
Let us begin with the first one. In the beginning there is one galaxy at the centre of each halo.
When two haloes merge, the central galaxies of the two haloes will sink towards the common centre of mass
owing to dynamical friction. That will take some time, so that the halo may temporarily be without a central galaxy.
The common centre of mass will be closer to the centre of mass of the more massive halo, so that
the central galaxy of the more massive halo will reach the centre of the new halo more rapidly,
not only because it is more massive and is therefore subject to stronger dynamical friction (computed from the standard 
differential equation in \citealp{binney_tremaine87}),
but also because it starts from a lower radial coordinate. 
Galaxy groups and clusters form because above a certain mass the dynamical friction time-scale for galaxy mergers becomes
longer than the time-scale on which dark matter haloes merge. When groups or clusters merge, the radial coordinates of 
satellite galaxies in the new halo are determined according to a prescription such that, if 
the satellites come from the main progenitor and this is almost as massive as the final halo,
their radial coordinates are nearly unperturbed. 
On the other, if the satellites come from a much less massive halo, their final radial coordinates are close to the virial radius of the new halo.
GalICS considers not only mergers at the centre of the halo due to dynamical friction, but also 
mergers due to encounters of galaxies at a non zero radial coordinate (collisions between satellite
galaxies), but these events are a small correction to the predominant merging activity with the central galaxy.

\subsubsection{Effect of mergers}
We now come to the second part of the merging model: how mergers transform galactic morphologies.
Bulge material remains bulge material and starburst material remains starbust material, while some of the disc material is removed from
the disc. As in the case of the bar instability, gas is transferred to the starburst and stars are transferred to the bulge.
The fraction of the disc mass transferred to the spheroidal component (the bulge and the starburst) depends on the mass ratio of the
merging galaxies. The exact formula for this fraction is in \citet{hatton_etal03}, but the idea is that when one 
of the two galaxies is much 
less massive than the other one (minor merger), most of the disc mass stays in the disc. On the contrary, if the two galaxies have
comparable masses (major merger), most of the disc mass is transferred to the spheroidal component.
The separation between a minor and a major merger is for a mass ratio of 1:3.

Devriendt et al. (in preparation) have used GalICS to investigate the relative importance 
of bars, minor mergers and major mergers in the formation of spheroids.
Elliptical galaxies acquire most of their mass through mergers, although disc instabilities are predominant in
forming the bulges of spiral and lenticular galaxies.
Minor mergers predominate over major mergers in all galaxy types.

\subsubsection{Tests of the model}
In Fig.~1, the predictions of the GalICS model for the mass function and the stellar ages of local spheroids are shown as solid lines.
These predictions are compared with a variety of data (shown as points with error bars, dashed and dotted lines). 

The mass function of spheroids is a key test of the model.
We have tried to make the comparison as robust as possible by using data derived with completely different methods.
The first is to start from the galaxy luminosity function decomposed by morphological
type, to assume a bulge fraction for each morphological type and finally to use a bulge
mass-to-light ratio to convert the spheroid luminosity function into a mass
function. We have applied this method to the 2Mass $K$ band data of
\citet{kochanek_etal01} and to the Sloan Digital Sky Survey $r^*$ band data of \citet{nakamura_etal02}, 
both with an aperture correction of -0.2$\,$mag
(see \citealp{shankar_etal04}, from where we have taken the bulge-to-total mass
ratios for early-type and late-type galaxies, and the $K$ band and $R$ band
mass-to-light ratios).
With these data, we have found the mass functions shown by the open diamonds
\citep{kochanek_etal01} and the dotted line \citep{nakamura_etal02} in Fig.~1.
The second way to compute the spheroid mass function is to perform a bulge to
disc decomposition on a complete galaxy sample, to determine a bulge mass
function and then to use a bulge mass-to-light ratio.
The open triangles show the result of this method with the $I$ band spheroid
luminosity function of \citet{benson_etal02}.
The third method does not pass through galaxy luminosities at all.
Instead, it uses the relation of bulge mass and velocity dispersion to convert
the Sloan early-type galaxies velocity dispersion distribution
(e.g. \citealp{cattaneo_bernardi03}) into a spheroid mass function.
The spheroid mass function computed through this method is shown by the dashed
line (which misses the bulges of spirals because it is derived from a sample of early-type galaxies).
The luminosity functions by \citet{kochanek_etal01} and \citet{nakamura_etal02}
give very similar results.
The mass functions obtained from \citet{benson_etal02} and
\citet{cattaneo_bernardi03} resemble each other very closely, too, while there is difference between the
first and the second group. However, the overall impression is that there is good agreement
between the model and the data.
In this comparison, we should have in mind that
the dark matter simulation cannot resolve haloes less massive than $1.6\times 10^{11}M_\odot$.
(haloes that contain less than $2\times 10^{10}M_\odot$ in baryons).
Therefore, a galaxy less massive than $2\times 10^{10}M_\odot$ is formally below the resolution limit.
In the case of Fig.~1, the importance of resolution effects is reduced by the fact that 
the fraction of the baryonic mass in bulges is small in low mass haloes.

The second panel of Fig.~1 shows the mean stellar ages of spheroids in GalICS (solid line) and in the
Sloan Digital Sky Survey (points with error bars). The small points show the scatter in GalICS
ages. The points with error bars come from \citet{cattaneo_bernardi03} using 
$M_{\rm bulge}\sim 10^{11}(\sigma/200{\rm\,km\,s}^{-1})^4$, where $\sigma$ is the stellar velocity dispersion of the spheroid.
The stellar ages of spheroids that we are comparing are not exactly the same thing because the model outputs are mean values weighted by stellar mass, while the mean values inferred from the Sloan Digital Sky Survey are weighted by luminosity, but the difference should not be large for the 
$r$-band, especially for massive galaxies with old stellar populations.
The agreement is acceptable but not so good as that found for the spheroid mass function.
In the cosmology used for this paper the Universe is 13.7$\,$Gyr old.
Lookback times of 7, 10 and 11$\,$Gyr correspond to redshifts of 1, 1.7 and 2.3, respectively.
Low mass ellipticals are older in the model than in the data because at low masses disc galaxies contaminate the early-type 
sample, but the real problem is at large masses. GalICS shows that the typical redshift of formation of the stars in a galaxy 
with $M_{\rm bulge}\simeq 2.5\times 10^{11}M_\odot$ is of $z\sim 1.1$ whereas the data suggest a value closer to $z=1.7$.
This discrepancy is a sign that the formation of spheroids in semi-analytic models of galaxy formation is an open problem
and indicates the need of more work on this topic.

In Fig.~2 we show the cosmic evolution of the galaxy merging rate by plotting total masses (stars and cold gas) and gas masses 
for all mergers that have occurred in 4 time spans of 100 Myr each at redshifts of about 2, 1, 0.5 and 0. 
This figure contains more information than
a plot of the merging rate as a function of redshift for galaxies above a mass threshold.
The decrease of the merging rate as a function of redshift is not dramatic because the larger
intergalactic distance at lower density of is compensated by a larger galaxy number, especially at high galaxy masses.
On the other hand, low redshift mergers have a tendency to a lower gas fraction.

\subsection{Galaxy luminosities}

Stellar spectral energy distributions (SEDs) in the optical and near infrared are computed by convolving the 
star formation history of each component with the SEDs derived from the STARDUST 
\citep{devriendt_etal99} stellar population synthesis model. STARDUST calculates the SEDs associated with a 
single burst of star formation at time intervals that go from 10\,Myr to 50\,Gyr. These SEDs assume a
Kennicutt stellar initial mass function and depend on the metallicity of the stellar population.
Dust absorption is computed with an extinction law depending on the mean hydrogen column density 
and the gas metallicity. 
The column density depends on the mass and the geometry of the gas distribution. 
GalICS uses two models: a spheroidal distribution for bulges and starbursts and a uniform slab for discs. 
In both cases, stars and dust have the same space distribution.
For each disc, GalICS picks a random inclination angle and computes the extinction for that value.
All the radiation that is absorbed is re-emitted in the infrared by four dust components: big carbonaceous 
grains, small carbonaceous grains, silicates and polycyclic aromatic hydrocarbons. 
The infrared colour-luminosity relation 
observed in IRAS galaxies determines how the absorbed power is distributed among the four components. 

\section{The structure of the AGNICS model}
 
\subsection{Black hole growth model}

The most recent data suggest that most bulges contain a supermassive black hole (see the Introduction).
We do not know how these black holes formed, although \citet{rees84} identified a number of astrophysical
paths that are likely to result in the formation of a supermassive black hole.
The main problem is how gas concentrates from a scale of $\sim100\,$pc down to $\sim1\,$pc.
We do not try to model the complicated physics of how gas flows from the galaxy into the
accretion disc and then from the accretion disc into the supermassive black hole.
Instead, we accept as a fact that whenever there is a bulge or a starburst, there is a supermassive black hole.

The model for black hole growth is based on three ingredients: the initial black hole mass,
the model for black hole accretion and the model for black hole coalescence.

\subsubsection{Seed mass}
The initial black hole mass is important in models where the black hole mass determines the
maximum accretion rate. If the accretion rate cannot exceed the Eddington limit (discussed in Section 3.2), 
then a larger initial mass allows the black hole to grow more rapidly. 
If we do not limit the accretion rate to the Eddington limit, then the value of the initial black hole mass is irrelevant 
(at least as long as the initial black hole mass is $M_\bullet\lsim 10^5M_\odot$).

\subsubsection{Accretion rate}
The model for black hole accretion is the fundamental element that distinguishes between different models 
in the literature and also
between different implementation of the AGNICS model. 
There are two families of models: those in which black holes grow through the accretion of cold gas 
(e.g. \citealp{cattaneo_etal99,kauffmann_haehnelt00,enoki_etal03}), in which case we expect a relation between 
black hole growth and star formation, and those in which black hole grow through spherical \citet{bondi52} 
accretion of hot gas (e.g. \citealp{nulsen_fabian00}).
In this paper we only fuel AGN with cold gas.
We tried to develop models of the second type but we were not able to do it
without overpredicting the number density of bright quasars at low redshift.
The reason is that, at low redshift, potential wells are deeper and haloes are less dense, with
the consequence that shock heating is more effective, the cooling time is longer and 
a higher baryonic fraction is in hot gas.

Within the family of models in which black hole accretion is fuelled with cold gas and triggered by mergers, we can
identify different prescriptions for the gas mass accreted by the black hole.
\citet{cattaneo_etal99} assumed that at each major merger the black hole accretes a mass 
$\Delta M_\bullet\propto (1+z)^\eta\Delta M_{\rm *burst}$, where $\Delta M_{\rm *burst}$ 
is the stellar mass formed by the merger
(in practice the mass of the gas in the discs of the merging galaxies) 
and $\eta$ is a free parameter determined to reproduce the
cosmic growth of the comoving mass density in supermassive black holes (a reasonable agreement with the data required 
that $\eta\sim 2$).
\citet{kauffmann_haehnelt00} assumed that on average, at each major merger, the black
hole accretes a mass 
\begin{equation}
\label{kh}
\Delta M_\bullet={f_{\rm BH}M_{\rm cold}\over1+(280{\rm\,km\,s}^{-1}/V_{\rm c})^2},
\end{equation}
where $M_{\rm cold}$ is the mass of cold gas in the merging galaxies, $V_{\rm c}$ is the host halo
circular velocity and $f_{\rm BH}$ is a free parameter determined from the
black hole-to-bulge mass relation.
The accretion was distributed over time with the law:
\begin{equation}
\dot{M}_\bullet={\Delta M_\bullet\over t_{\rm accr}}{\rm\,exp}\left(-{t\over t_{\rm accr}}\right),
\end{equation}
where $t_{\rm accr}(z)$ is a free parameter determined by fitting the quasar luminosity function.
\citet{enoki_etal03} used a very similar model with 
\begin{equation}
\label{enoki}
\Delta M_\bullet=f_{\rm BH}\Delta M_{*,{\rm burst}}=
{f_{\rm BH}M_{\rm cold}\over 0.75+(280{\rm\,km\,s}^{-1}/V_{\rm c})^{2.5}}
\end{equation}
where $\Delta M_{\rm *burst}$ is the total mass of the stars formed in the
starburst that accompanies the merger.

In AGNICS we integrate a differential equation for the black hole mass.
In this paper we only consider models of the form:
\begin{equation}
\label{bar}
\dot{M}_\bullet=k \dot{M}_{\rm *burst}.
\end{equation}
$\dot{M}_{\rm *burst}$ is the star formation rate in the starburst component.
Black holes in starbursts are active while
black holes in galaxies that do not contain a starburst are quiescent.
The accretion efficiency $k$ can depend on an large number of parameters
(morphologies of the merging galaxies, gas fraction, structure of the intergalactic medium, 
effect of AGN and supernova feedback,
impact parameter, inclination of the orbital plane with
respect to the discs of the merging galaxies, corotation or counterrotation of the two discs, 
black hole spin, etc.).
In practice, we only consider very simple cases. In this paper we discuss four of them
(Table~1 summarizes all the models considered in this paper):
\begin{itemize}
\item $k=$constant: there is a one-to-one relation between black hole growth and starbursts.
This is what we called the basic model and corresponds to model A.
\item $k=$constant, $\dot{M}_\bullet\le\dot{M}_{\rm Edd}$ (model B).
\item $k\propto\rho_{\rm cold}^\zeta$ where $\rho_{\rm cold}$ is the density of the gas in the starburst component.
This is the reference model, which gives the best fit to the data and corresponds to models C, D and E. 
\item $k\propto\rho_{\rm cold}^\zeta/[1+(280{\rm\,km\,s}^{-1}/V_{\rm c})^2]$ 
to suppress black hole growth in small haloes (model F).
\end{itemize}

These recipes describe a mean accretion rate: the approach model oversimplifies the physics of individual
objects, but can be expected to be meaningful in a statistical sense, when it is used to compute mean values in a cosmological volume.
However, neglecting the dependence of $k$ on other parameters will lead to underestimate the scatter in the properties of black
holes and AGN. The scatter in the results of the simulations can only be
meaningful as a lower limit to the scatter predicted by the model.
Another remark: Eq.~(\ref{bar}) implies that the AGN duty cycle coincides with the entire duration of the starburst.
This is a simplification because black hole accretion is likely to be highly time-dependent, as it is shown by AGN variability.

\subsubsection{Black hole coalescence}
The third choice that one has to make to complete the black hole growth model is what happens when two galaxies with
black holes merge. We assume that the black holes immediately coalesce prior to any gas accretion.
This is not a good description of reality if the active phase starts before the black holes have merged 
(it is not difficult to find images of quasar hosts with double nuclei).
It is also possible that a third black hole from a second merger reaches the centre before
the two black holes from the first merger have had time to coalesce. In this case, the three-body interaction 
may result in the ejection of one black hole from the galaxy
\citep{begelman_etal80,hut_rees92,volonteri_etal03a}.
Nevertheless, the simple assumption of instantaneous coalescence
is the most natural within the assumption of instantaneous 
morphological transformation at the time of merging used in semi-analytic models.

\subsection{Luminosity and spectral energy distribution of AGN}

The bolometric luminosity of a quasar is 
\begin{equation}
\label{bol}
L_{\rm bol}={\epsilon_{\rm rad}\over 1-\epsilon_{\rm rad}}\dot{M}_\bullet{\rm c}^2,
\end{equation} 
where $\epsilon_{\rm rad}\sim 0.1$ is the radiative efficiency of black hole
accretion. \citet{bardeen70} explains the physics of this value.

The Eddington luminosity $L_{\rm Edd}$ is the critical luminosity above which the radiation pressure is stronger than 
the gravitational attraction. 
The justification for requiring $L_{\rm bol}\le L_{\rm Edd}$
is that, as soon as the Eddington luminosity is exceeded, radiation pressure pushes the gas 
outwards and brings the accretion rate back to sub-Eddington values. 
However, there are physical situations in which this limit can be exceeded, e.g. if the emission is beamed 
into a narrow cone or if photons are trapped into a radiatively inefficient, optically thick flow and advected into the event 
horizon \citep{begelman78}. In the latter case it is possible to have $\dot{M}_\bullet>\dot{M}_{\rm Edd}$
without requiring  that $L_{\rm bol}> L_{\rm Edd}$ (here $\dot{M}_{\rm Edd}$ is the black hole accretion rate for which
Eq.~\ref{bol} gives $L_{\rm bol}=L_{\rm Edd}$).
Imposing the condition $\dot{M}_\bullet\le\dot{M}_{\rm Edd}$
introduces a characteristic time-scale for the growth of the black hole.
This time-scale is linked to the Salpeter time $t_{\rm S}=4\times 10^8\,$yr,
the time in which a black hole radiating at the Eddington limit releases its entire mass energy
$M_\bullet{\rm c}^2$. Since a black hole cannot radiate its entire mass
energy, the accretion time-scale is limited to $t_{\rm accr}=\epsilon_{\rm rad}t_{\rm S}$
and the black hole mass cannot grow faster than exp$(-t/t_{\rm accr})$.

\citet{mirabel_rodriguez99} have shown that some microquasars in the Milky Way are accreting well above
the Eddington limit.     
In this paper we consider different models (see Table 1): the default assumption is that nothing stops black holes
from accreting at super-Eddington rates, but the luminosity cannot exceed the Eddington limit.
This is the same assumption of \citet{kauffmann_haehnelt00}.

The standard reference in the literature for the quasar SED is the
\citet{elvis_etal94} median SED inferred from radio,
sub-millimetre, infrared, optical, ultraviolet and X-ray observations of a sample of optically selected quasars. 
With this template, the blue magnitude of an optical quasar is
\begin{equation}
M_B=-23.64-2.5{\rm\,Log}(L_{\rm bol}/10^{46}{\rm\,erg\,s}^{-1})
\end{equation}
and the colours are: $U-B=-0.911$, $B-V=0.092$, $V-R=0.320$ and $R-K=3.287$.
The \citet{elvis_etal94} SED gives a bolometric correction of $\simeq 32$ for the 2-10$\,$keV X-ray band. 
\citet{marconi_etal04} have shown that a luminosity-dependent bolometric correction provides a more adequate fit to the
X-ray data and it is their model that we use to compute X-ray luminosities.

\subsection{Obscured AGN}
\begin{figure}
\label{bhvsigma} 
\centerline{\psfig{figure=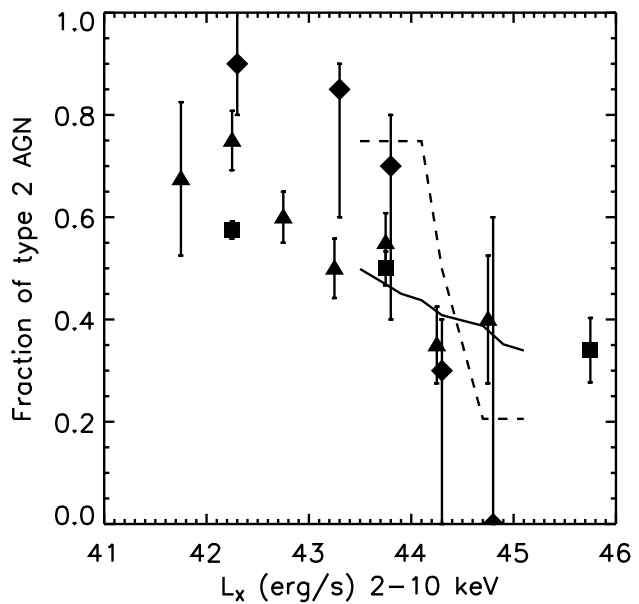,width=0.48\textwidth,angle=0}}
\caption{\small The fraction of X-ray selected AGN that do not have an AGN optical/UV continuum or broad emission lines. 
The squares are from \citet{ueda_etal03}, the triangles are from \citet{szokoly_etal04} and
the diamonds are from \citet{barger_etal05}. The solid line (extinction law I) and the dashed line 
(extinction law II) are two models used to fit these data.}
\end{figure}

Obscured AGN are AGN where the optical/UV continuum and the broad line spectrum are not observable.
Unified models state that unobscured and obscured AGN are intrinsically the same type of systems, made of a central black hole
surrounded by its accretion disc and, more further out, by a dusty torus.
In obscured (also called type 2) AGN, 
the system is observed from the equatorial plane of the torus and the optical/UV continuum is not observable because
it is absorbed by dust on the line of sight.
Obscured AGN can be detected in hard X-rays, at optical wavelengths (through scattered polarised light and strong narrow line emission),
by using mid-infrared spectroscopy and with radio observations (see the review article by Risaliti \& Elvis 2004).
\citet{ueda_etal03,szokoly_etal04,barger_etal05} used X-ray selected sample to estimate the fraction of obscured AGN
and found that the result is strongly luminosity-dependent (Fig.~3). Most bright quasars are unobscured, while $>75\%$ low luminosity AGN
are obscured. These findings are consistent with those by \citet{sazonov_etal04}, who studied 
the combined cosmic infrared and X-ray background and inferred an obscured/unobscured ratio of a factor of three.
Optical extinction is therefore an important effect and can shift the peak of the cosmic accretion history of supermassive black holes
to lower redshifts (e.g. \citealp{cattaneo_bernardi03,steffen_etal03}).

The \citet{elvis_etal94} SED is a phenomenological template for unobscured (type 1) quasars.
It contains three components: the big blue bump (due to thermal emission from the accretion disc), 
the infrared bump (due to thermal emission from the inner edge of the dusty torus) 
and the X-ray bump (due to Comptonization of accretion disc light). 
We cannot treat AGN extinction as we do with the absorption of stellar light by dust, where we assume a 
covering factor of $\sim 1$, because that grossly underpredicts the number of AGN by extinguishing the entire quasar
population with $A_V\sim 5$.
Instead, we assume that for a given bolometric luminosity there is an obscuration probability $f_{\rm obs}(L_X)$, 
which is determined phenomenologically from the data in Fig.~3. 
This is the same as to assume that the torus has a luminosity-dependent opening angle.
When an AGN is viewed from the equatorial plane of the torus we still see the infrared thermal emission from the torus and the
X-ray emission from the central engine (assuming that we can neglect self-absorption
at infrared wavelengths and that the torus is not Compton-thick).

The most delicate point is the choice of the function $f_{\rm obs}$. Fig.~3 shows the two model functions used in this paper.
The first one, shown by the dashed line, fits the data \citet{ueda_etal03} and \citet{szokoly_etal04}, 
while the second one,
shown by the solid line, is closer to the new data by \citet{barger_etal05}.
The two lines cover only a small fraction of the luminosity range because AGN with
$L_X\lsim 3\times 10^{43}{\rm\,erg\,s}^{-1}$ are too faint to contribute to the luminosity function of optical quasars and we do not want to extrapolate the curves beyond the range to which we are applying them
as a phenomenological model for the extinction of quasars by dust.

\section{Black hole masses and the redshift evolution of the quasar population}
\begin{table*}
\begin{tabular}{l|c|c|c|c}
\hline
\hline 
Model     &  $k$               & Eddington limit  & $\epsilon_{\rm rad}$ & Extinction \\
\hline
A (the basic model)  & 0.0017              & No    &    0.26     &  II \\
B    & 0.0017              & $\dot{M}_\bullet\le\dot{M}_{\rm Edd}$   &   -   & - \\
C (the reference model)  &$0.0012M_{11}^{0.5}r_{\rm burst}^{-1.5}$ & $L_{\rm bol}\le L_{\rm Edd}$     &   0.1 & II \\
D       &$0.0012M_{11}^{0.5}r_{\rm burst}^{-1.5}$ & $L_{\rm bol}\le L_{\rm Edd}$     &   0.1 & I \\
E       &$0.0012M_{11}^{0.5}r_{\rm burst}^{-1.5}$ & No     &   0.08 & II \\
F    &$0.0012M_{11}^{0.5}r_{\rm burst}^{-1.5}/[1+(280{\rm\,km\,s}^{-1}/V_{\rm c})^2]$ &  $L_{\rm bol}\le L_{\rm Edd}$    &   0.1 & I \\

\hline
\hline
\end{tabular}
\caption{Models used to produce the figures of the paper. In the most general case, the black hole accretion 
efficiency $k$ (Eq.~\ref{bar}) depends on the mass of the gas in the starburst $M_{11}$ (in units of $10^{11}M_\odot$), on
the starburst radius $r_{\rm burst}$ (in kpc) and on the halo circular velocity $V_{280}$ (in units of
$280{\rm\,km\,s}^{-1}$). We have considered three models for the physics of the Eddington limit.
In models A and E, the Eddington limit is completely ignored. In model B, black holes are constrained to accrete at 
sub-Eddington rates. In models C, D and F, black holes can grow at super-Eddington rates but their luminosities cannot
exceed the Eddington limit.
The radiative efficiency $\epsilon_{\rm rad}$ is the one used in Eq.~(\ref{bol}) to convert the black hole accretion rate
into a bolometric luminosity.
The extinction laws I and II refer to the solid line and the dash line in Fig.~3, respectively.}
\end{table*}
\begin{figure*}
\noindent
\begin{minipage}{8.6cm}
  \centerline{\hbox{
      \psfig{figure=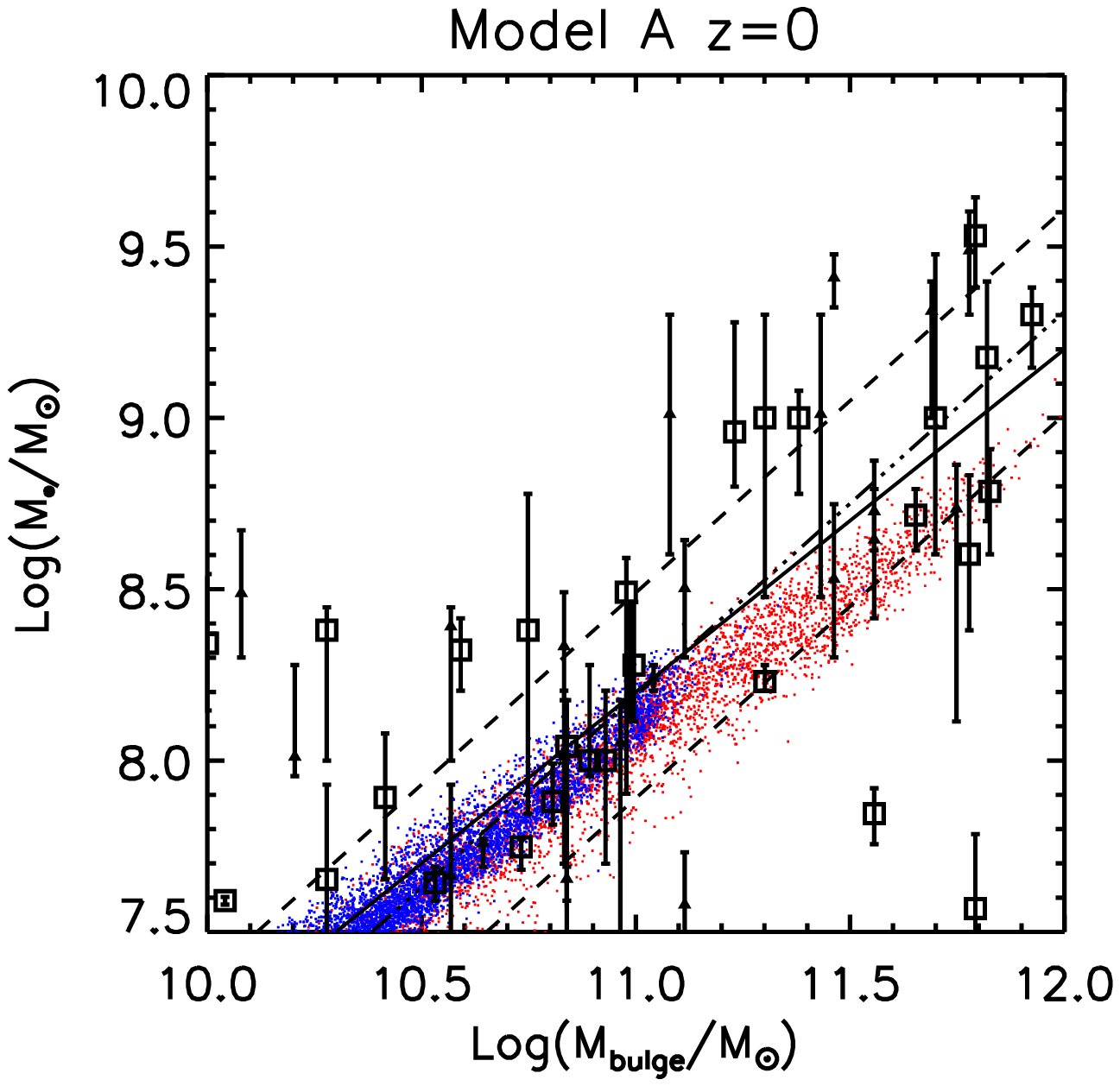,height=8.6cm,angle=0}
  }}
\end{minipage}\    \
\begin{minipage}{8.6cm}
  \centerline{\hbox{
      \psfig{figure=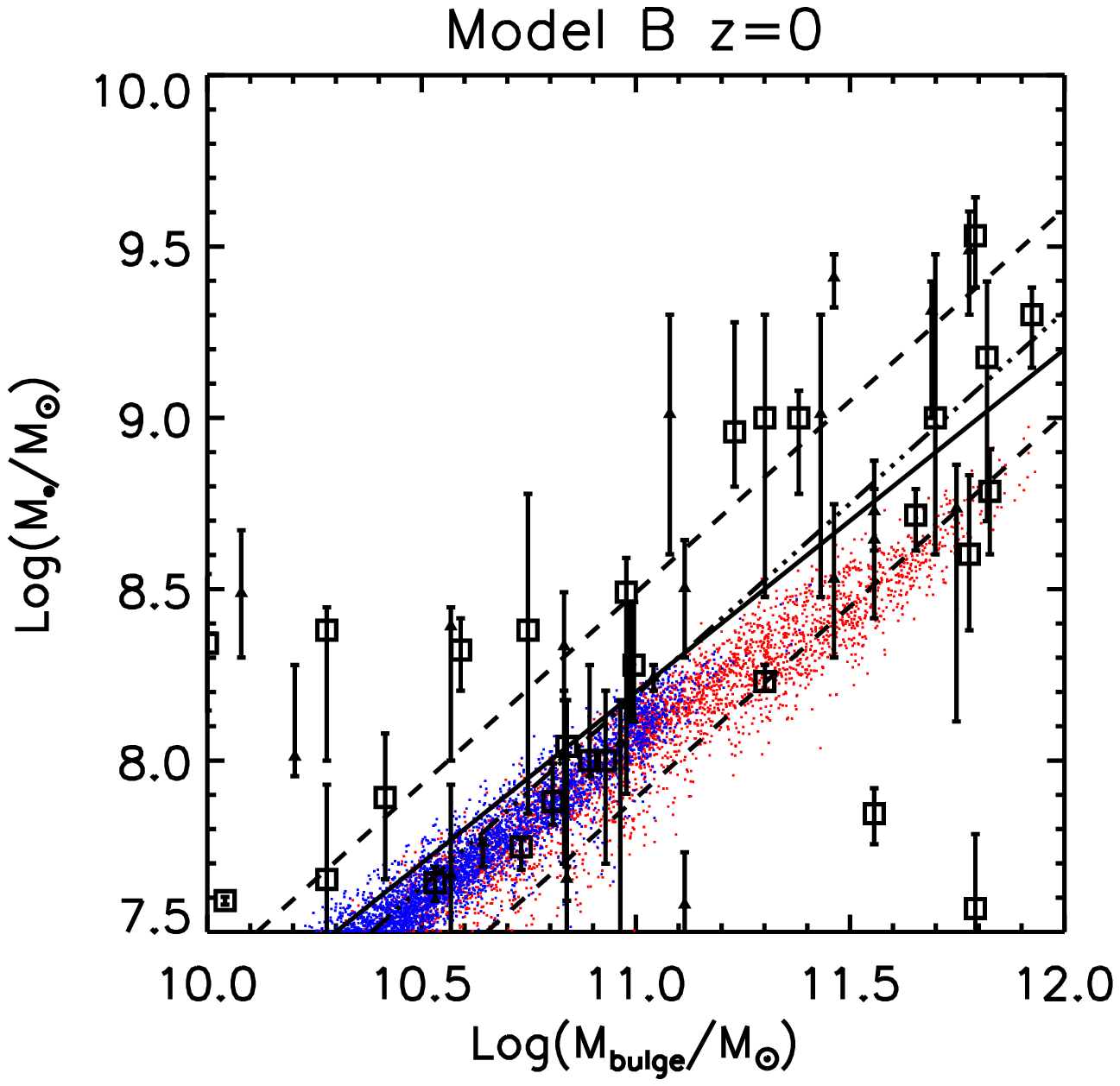,height=8.6cm,angle=0}
  }}
\end{minipage}\    \
\caption{The simulated black  hole mass - bulge mass relation in a model in which the black hole accretion rate is
directly proportional to the star formation rate in the starburst component ($\dot{M}_\bullet=0.0017\dot{M}_{\rm *burst}$). 
The left and the right panel shows the model predictions respectively without and with the condition 
$\dot{M}_\bullet\le\dot{M}_{\rm Edd}$ (models A and B of Table~1). Model B is computed for a seed mass of
$M_\bullet=10^{-7}M_{\rm cold}$ where $M_{\rm cold}$ is the mass of the gas in the starburst component. 
The predictions made by AGNICS are shown as dots, blue for black holes in disc-dominated galaxies, red for black holes in 
bulge-dominated galaxies.
The open squares are the mass estimates by \citet{marconi_hunt03} and the filled triangles those by \citet{haering_rix04}.
The dashed-dotted lines corresponds to a 1.12 slope (the best fit to the data of \citealp{haering_rix04}) 
and the dashed lines show a scatter of $\pm 0.3\,$dex around this relation.
The solid line is the \citet{marconi_hunt03} $M_\bullet=0.002M_{\rm bulge}$ relation rescaled for $M_\bullet=0.0017M_{\rm bulge}$.
These lines and the data points are the same in both panels.}
\end{figure*}
We want to determine if a model in which the black hole accretion rate is related to the star formation rate in the proto-spheroid
(Eq.~\ref{bar}) can reproduce the masses of black holes in nearby galaxies 
and the cosmic evolution of the quasar population.

\subsection{The basic model}

The simplest scenario is that the black hole accretion rate is directly proportional to the star formation rate in the
starburst component (models A and B of Table~1).
This assumption gives a tight $M_\bullet$ - $M_{\rm bulge}$ relation consistent with a linear scaling. 
The left and the right panel in Fig.~4 present the case 
without and with the constraint $\dot{M}_\bullet\le\dot{M}_{\rm Edd}$, respectively.
The blue and the red points are the results of the simulations for disc-dominated (late-type) and bulge-dominated (early-type)
galaxies.
The open squares are the mass estimates by \citet{marconi_hunt03} and the filled triangles those by \citet{haering_rix04}.
The solid line shows the linear relation $M_\bullet=0.0017M_{\rm bulge}$.
These lines and the data points are the same in both panels.

The tight $M_\bullet\propto M_{\rm bulge}$ relation follows directly from our assumption.
The $\sim 0.1\,$dex scatter in the left panel (model A) is due to the presence of a
stellar component in the discs of the merging galaxies. 
In a merger, the masses of the two bulges add together, and so do the masses of the two black holes.
The gas in the discs of the merging galaxies contributes to the formation of
bulge stars and to the growth of the black hole in a fixed proportion.
However, the stars in the discs of the merging galaxies can only contribute to the
growth of the bulge mass, and as
the gas fraction is lower at high masses (Fig.~2), high mass mergers produce lower values of $M_\bullet/M_{\rm bulge}$
and a shallower $M_\bullet$ - $M_{\rm bulge}$ relation at high masses.
The model with $\dot{M}_\bullet\le\dot{M}_{\rm Edd}$ (model B) contains more scatter because
the Eddington limit introduces a characteristic time-scale for the growth of the black hole.
In some galaxies with a short star formation time-scale,
the starburst may be over before the black hole has had time to grow significantly.
Nevertheless, the similarity between the two panels of Fig.~4 indicates that the importance of this effect is limited.

\begin{figure*}
\noindent
\begin{minipage}{8.6cm}
  \centerline{\hbox{
      \psfig{figure=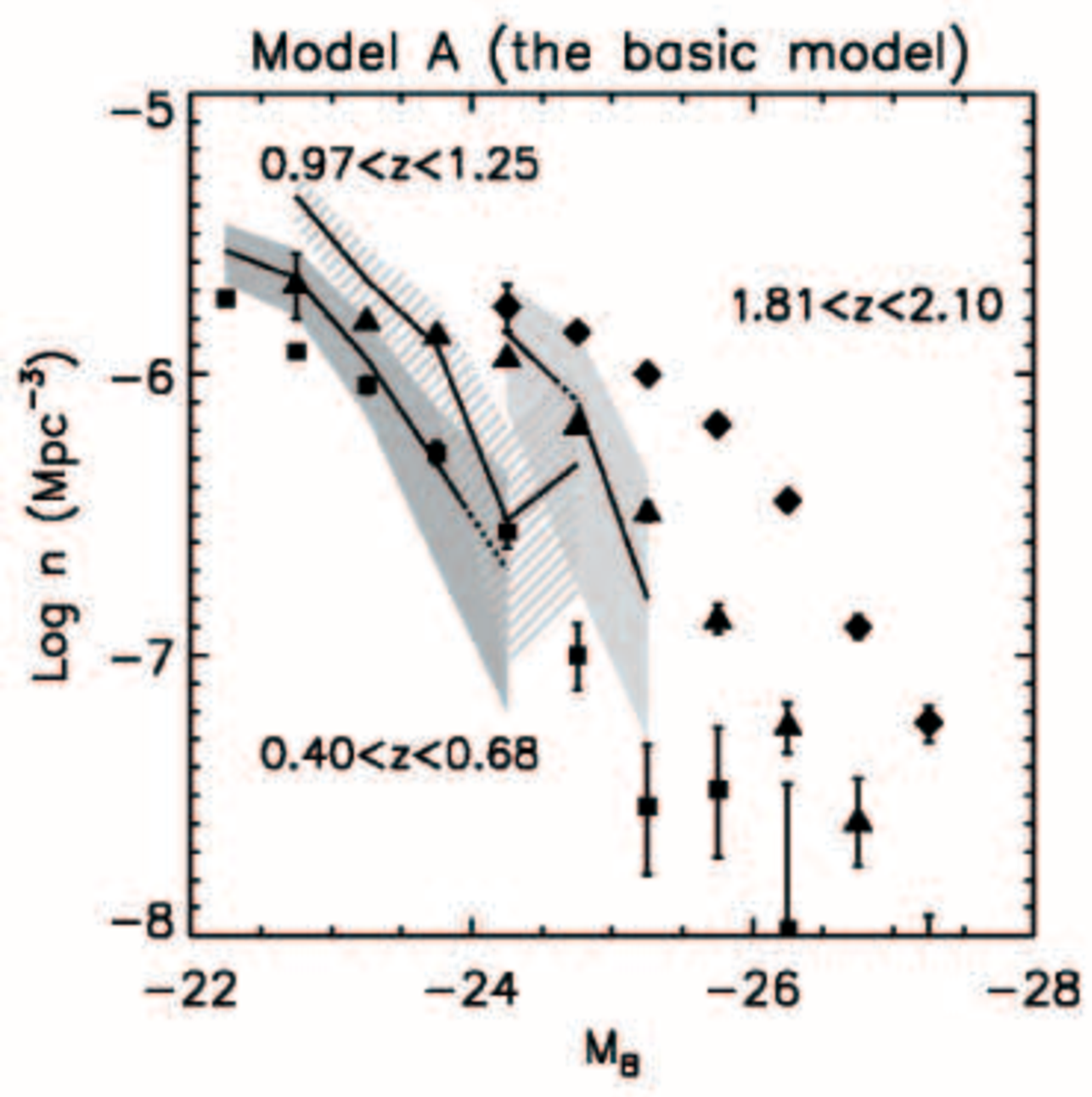,height=7.5cm,angle=0}
  }}
\end{minipage}\    \
\begin{minipage}{8.6cm}
  \centerline{\hbox{
      \psfig{figure=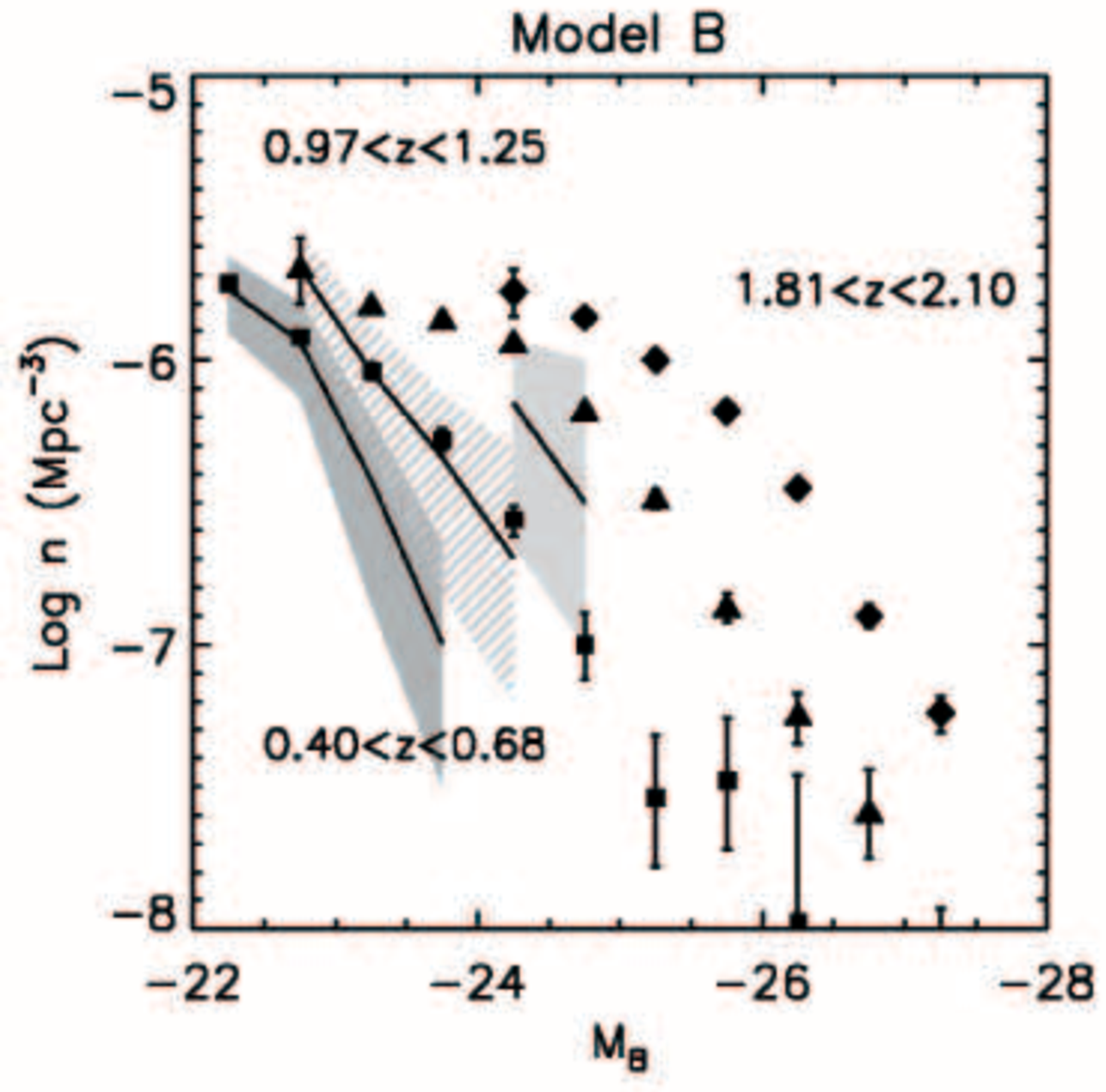,height=7.5cm,angle=0}
  }}
\end{minipage}\    \
\begin{minipage}{8.6cm}
  \centerline{\hbox{
      \psfig{figure=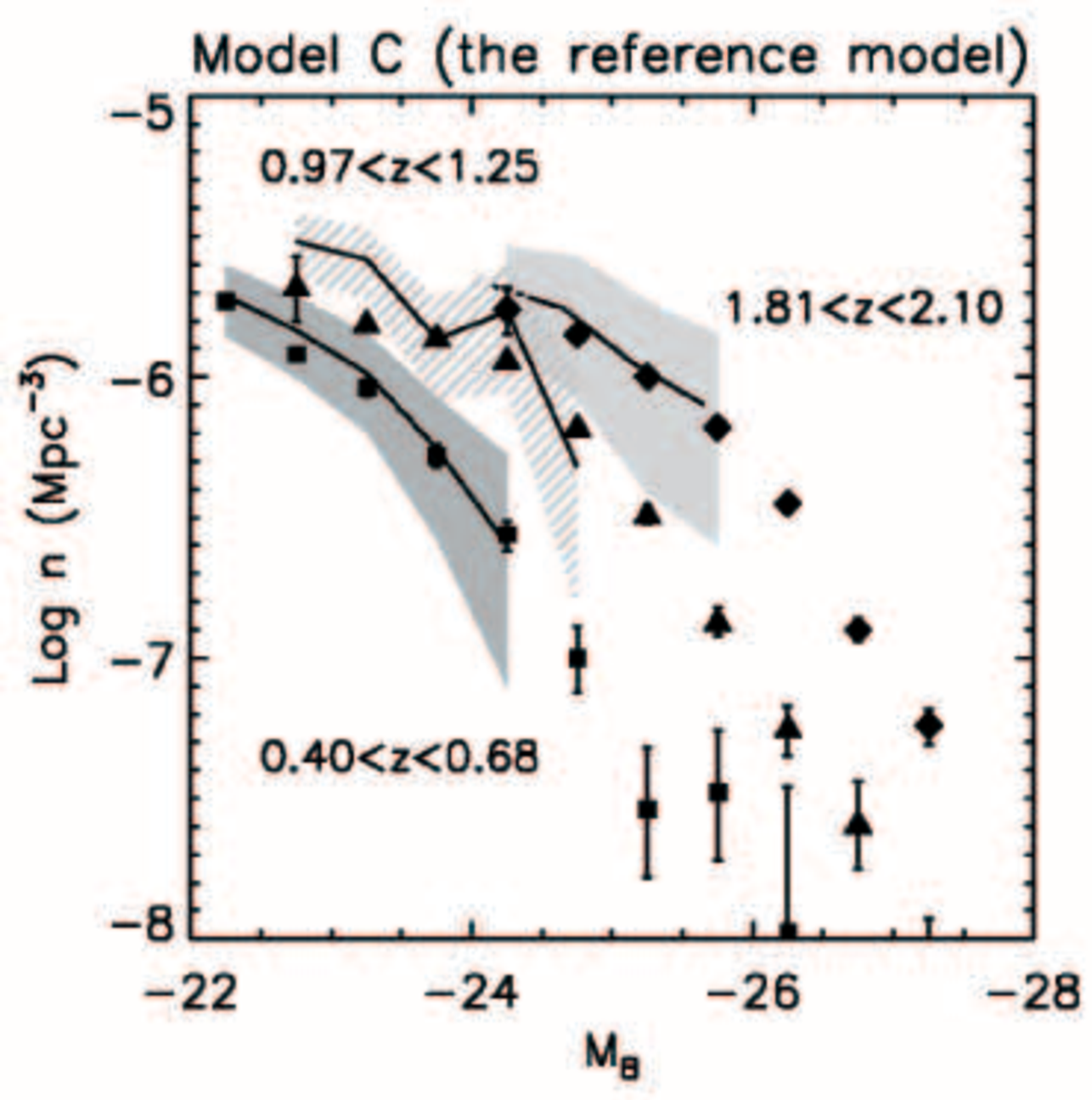,height=7.5cm,angle=0}
  }}
\end{minipage}\    \
\begin{minipage}{8.6cm}
  \centerline{\hbox{
      \psfig{figure=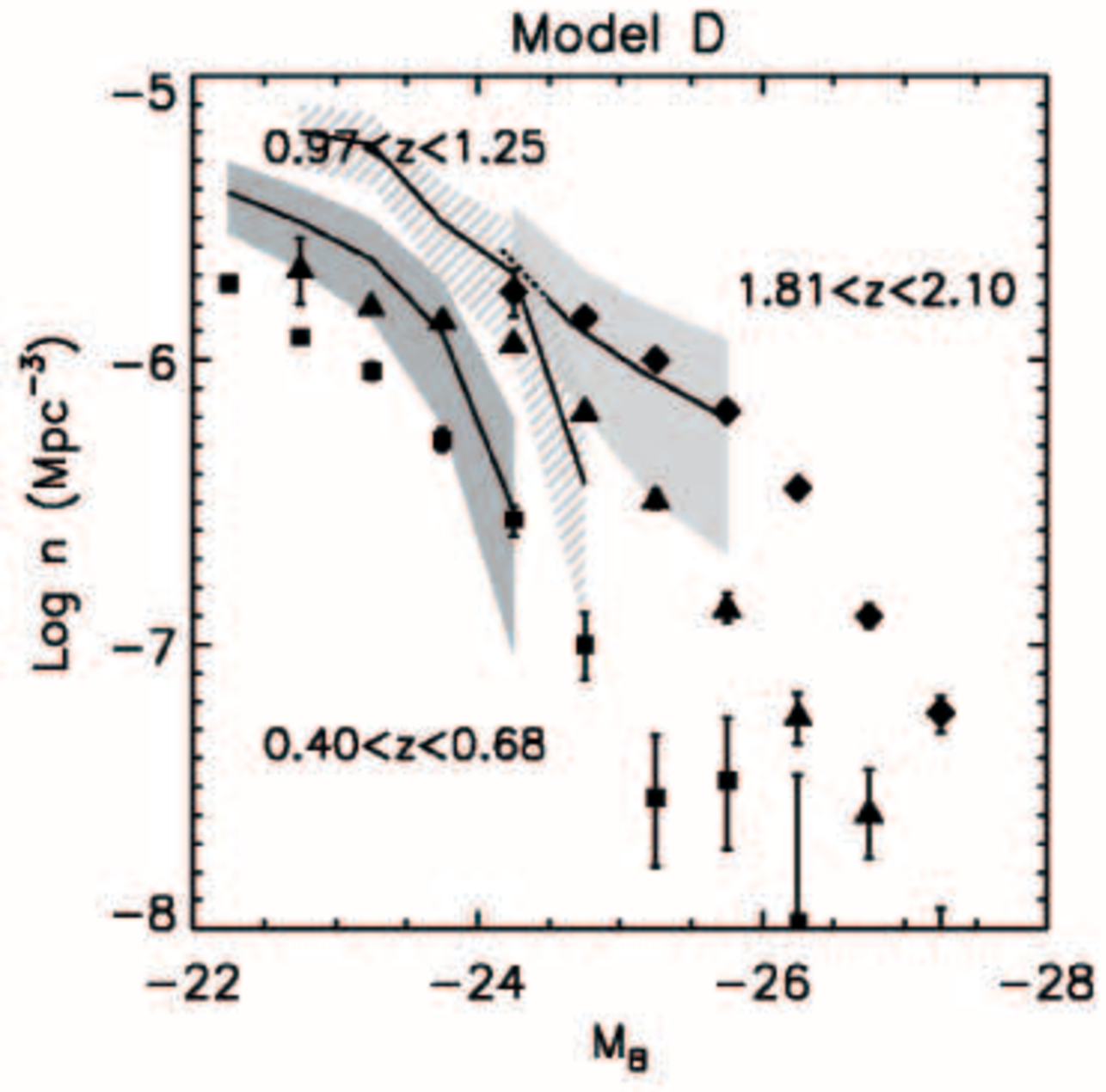,height=7.5cm,angle=0}
  }}
\end{minipage}\    \
\begin{minipage}{8.6cm}
  \centerline{\hbox{
      \psfig{figure=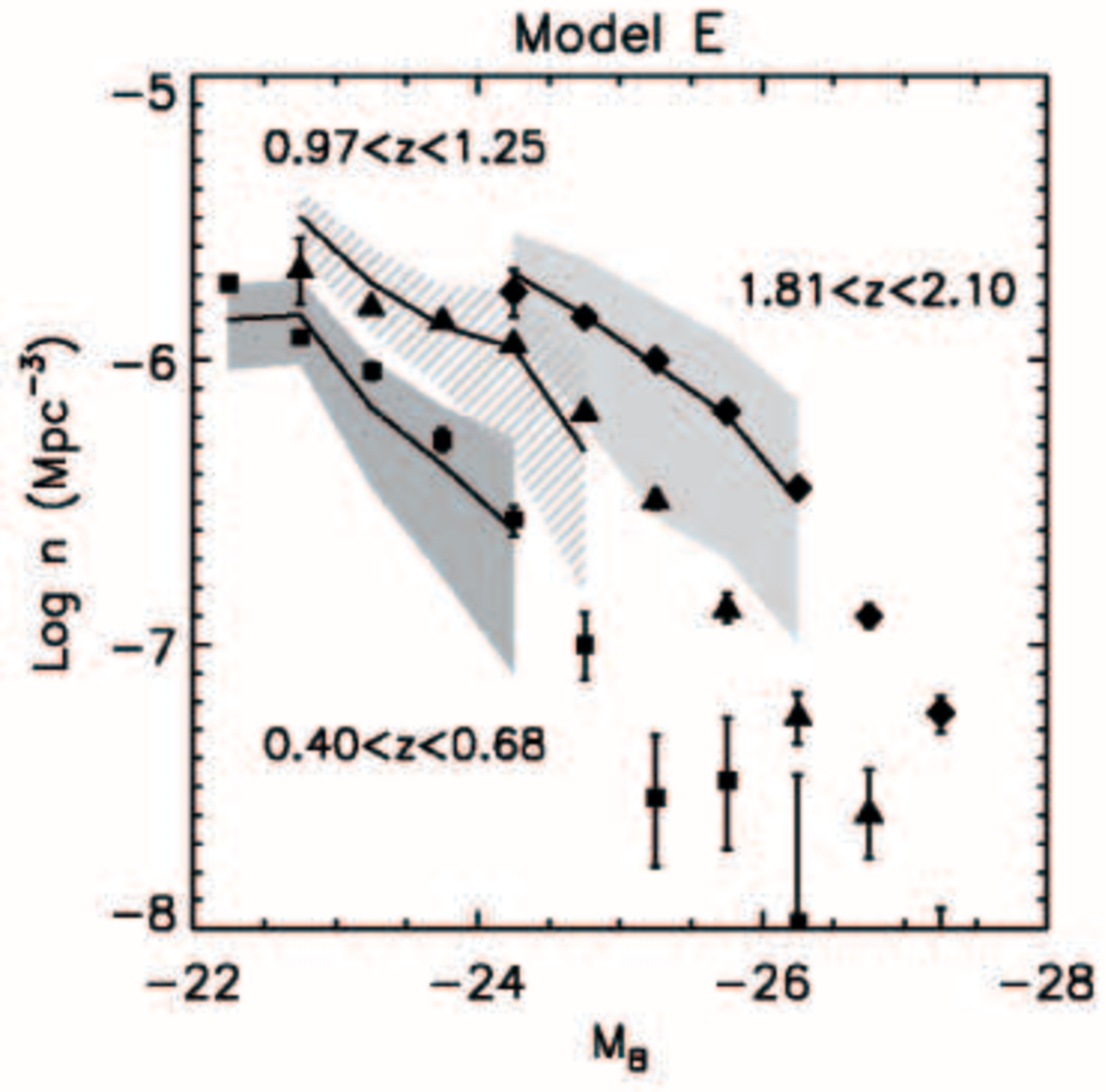,height=7.5cm,angle=0}
  }}
\end{minipage}\    \
\begin{minipage}{8.6cm}
  \centerline{\hbox{
      \psfig{figure=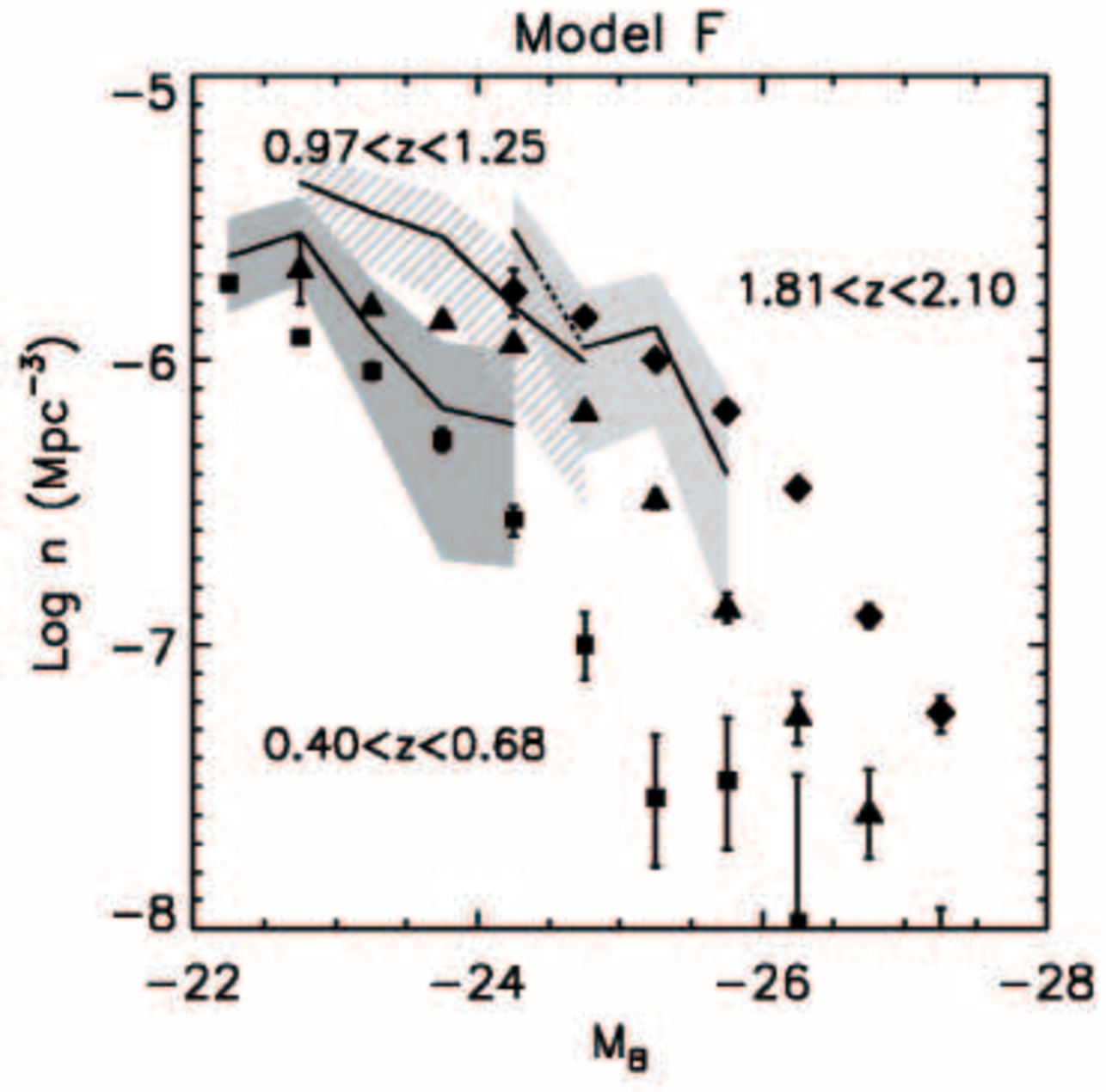,height=7.5cm,angle=0}
  }}
\end{minipage}\    \
\caption{\small The outputs of AGNICS (solid lines) for the six models listed in Table~1 compared with 
the 2dF quasar luminosity function of \citet{croom_etal04} (points with error bars).
The shaded areas show the error bars in the simulated luminosity functions.}
\end{figure*}
\begin{figure*}
\noindent
\begin{minipage}{8.6cm}
  \centerline{\hbox{
      \psfig{figure=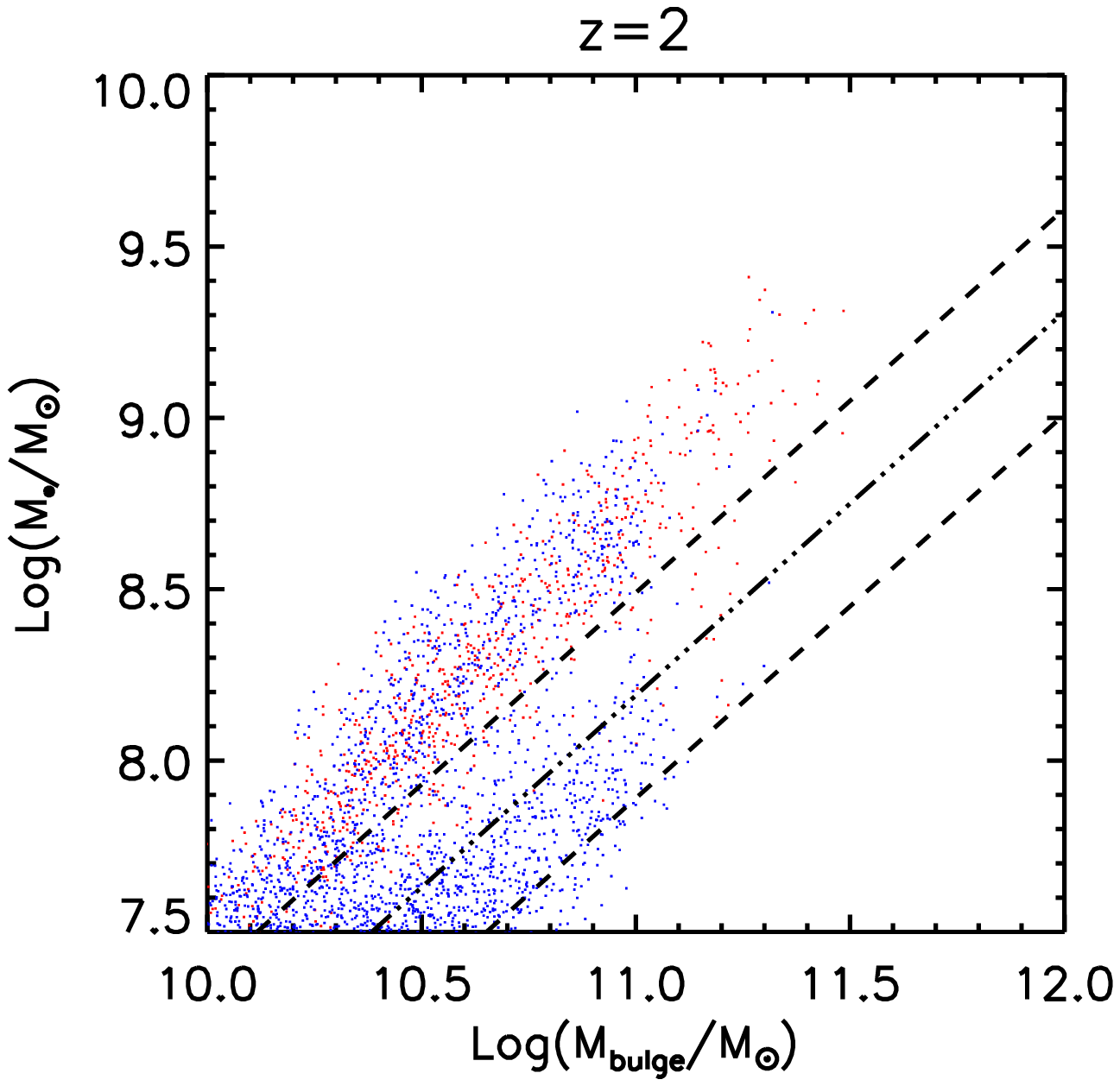,height=8.6cm,angle=0}
  }}
\end{minipage}\    \
\begin{minipage}{8.6cm}
  \centerline{\hbox{
      \psfig{figure=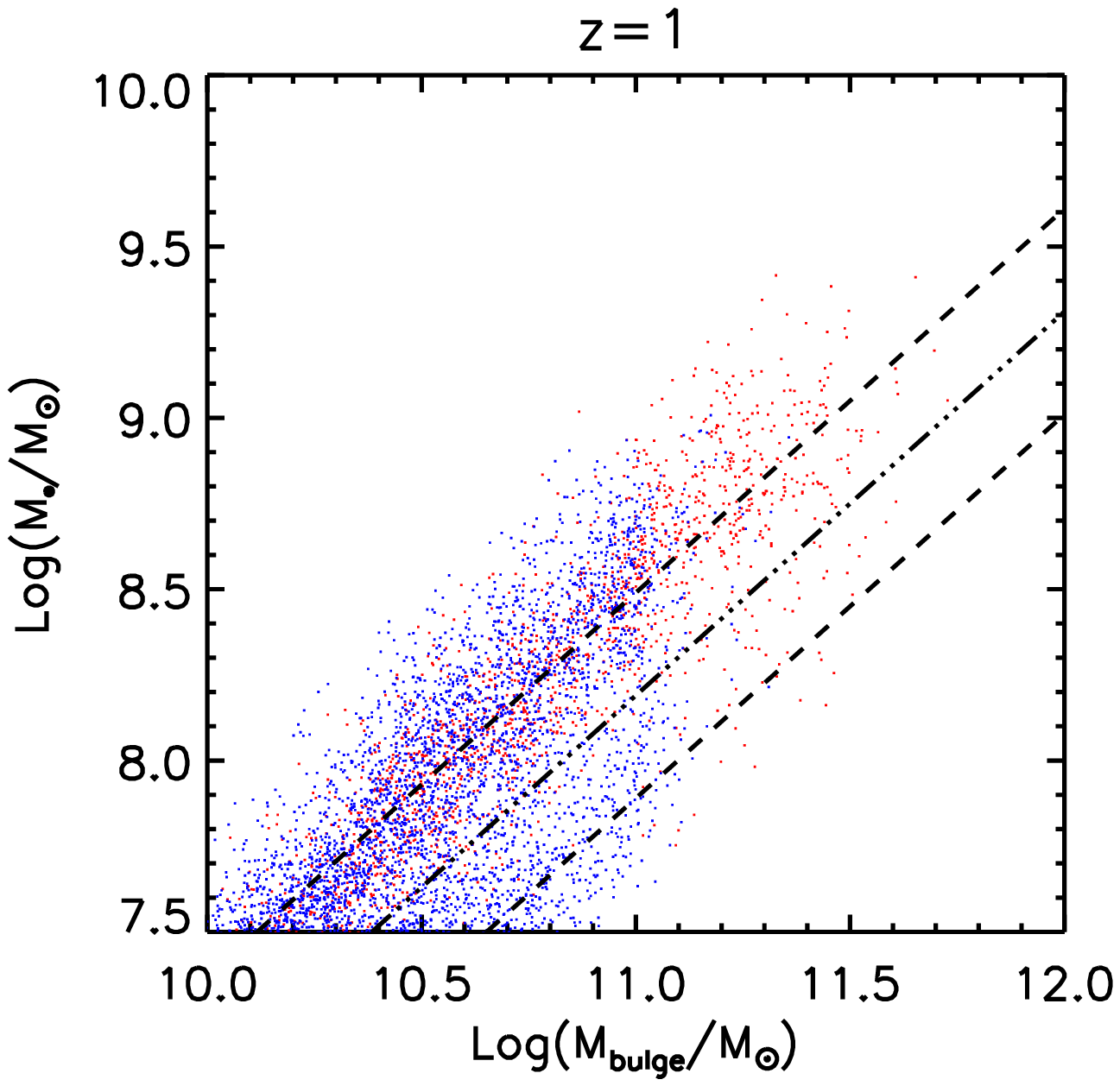,height=8.6cm,angle=0}
  }}
\end{minipage}\    \
\begin{minipage}{8.6cm}
  \centerline{\hbox{
      \psfig{figure=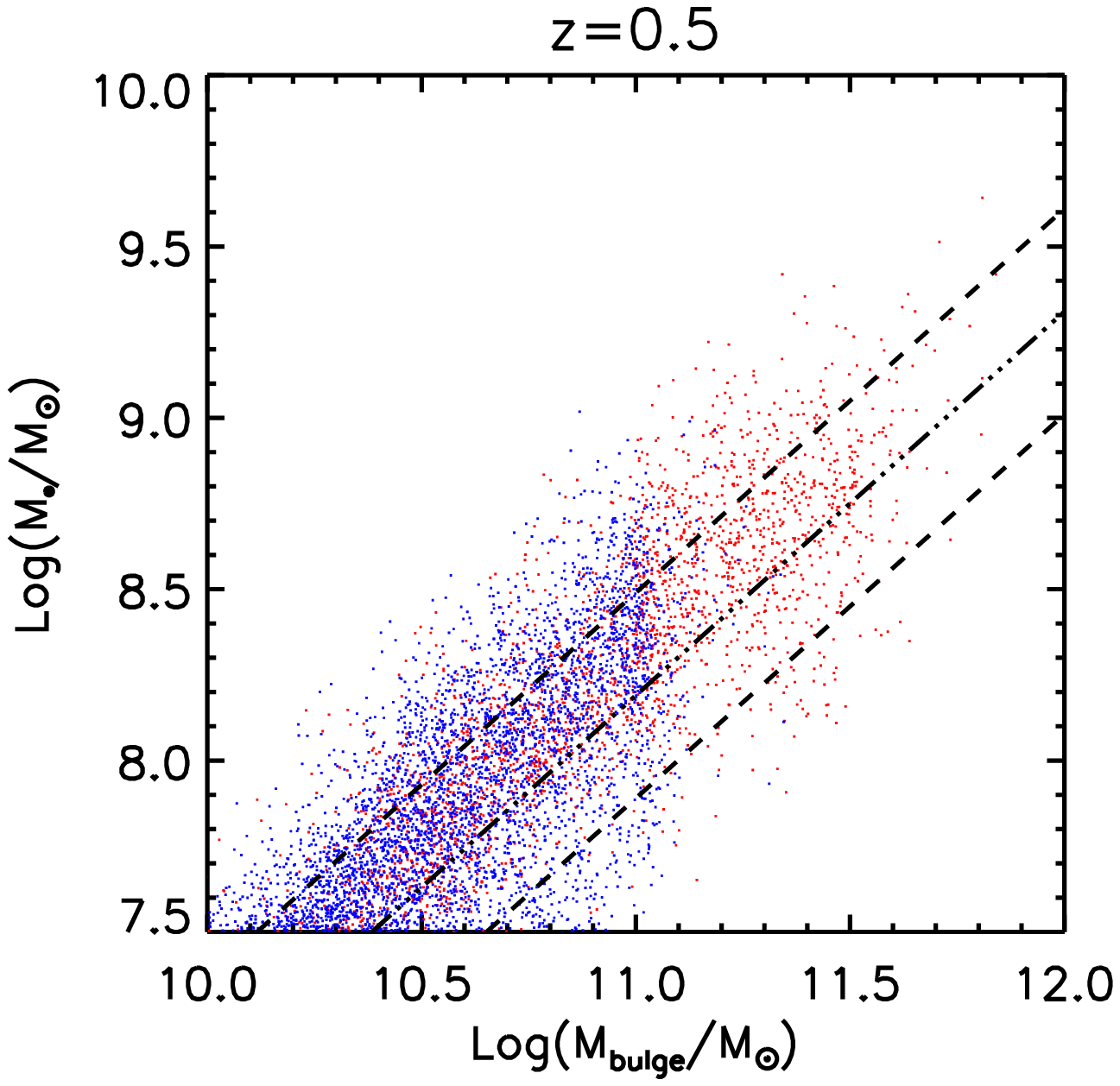,height=8.6cm,angle=0}
  }}
\end{minipage}\    \
\begin{minipage}{8.6cm}
  \centerline{\hbox{
      \psfig{figure=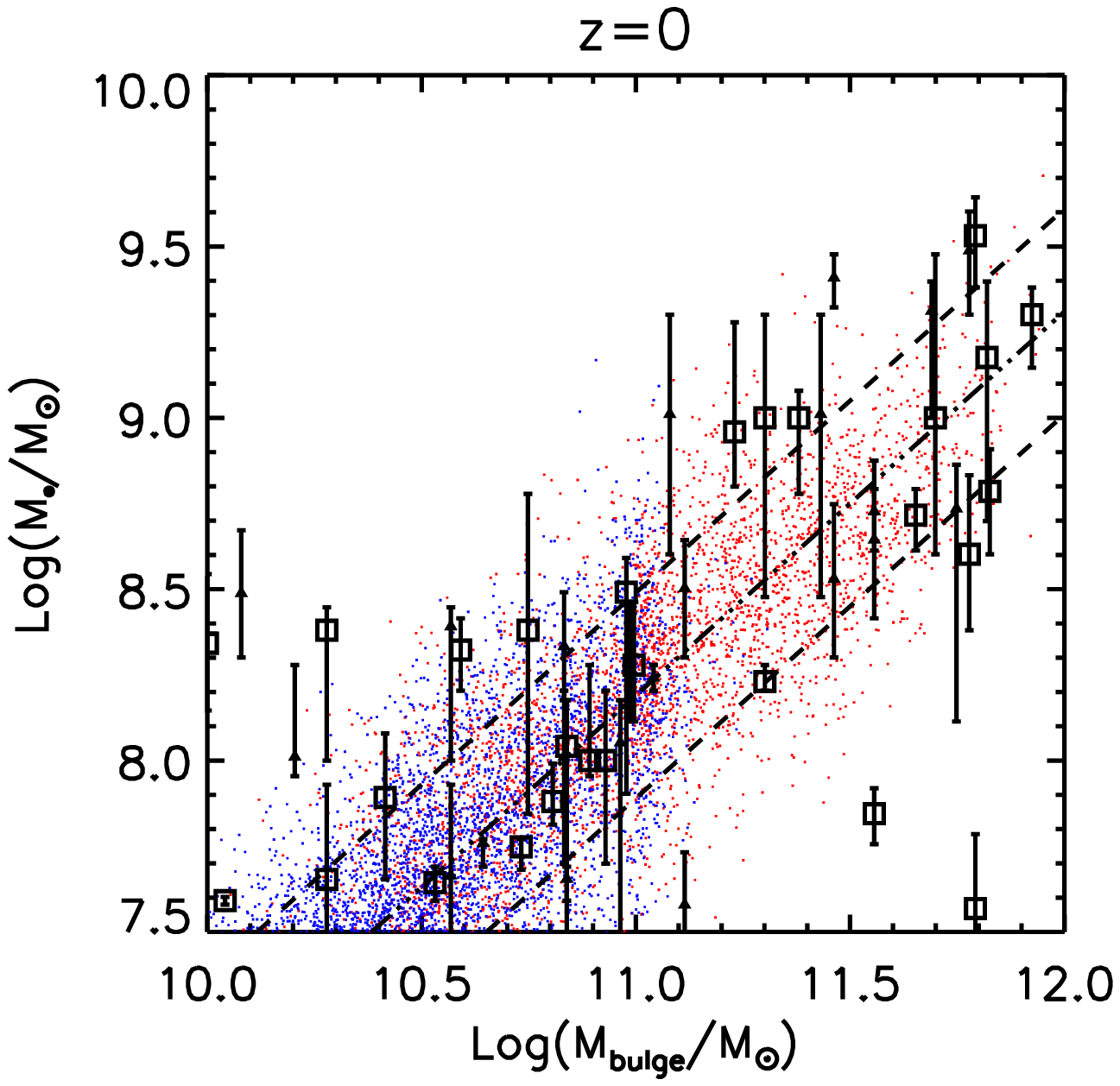,height=8.6cm,angle=0}
  }}
\end{minipage}\    \
\caption{\small The small points show the simulated black hole mass - bulge mass relation in a model in which 
$\dot{M}_\bullet=0.0012(M_{\rm cold}/10^{11}M_\odot)^{0.5}(r_{\rm burst}/{\rm kpc})^{-1.5}\dot{M}_{\rm *burst}$ 
(the reference model). 
Here $M_{\rm cold}$, $r_{\rm burst}$ and $\dot{M}_{\rm *burst}$ are the gas mass, 
the Hernquist radius and the star formation rate of the starburst component. 
The blue points are for black holes and bulges of late-type galaxies.
The red points are for black holes and bulges of early-type galaxies.
The open squares are the mass estimates by \citet{marconi_hunt03} and the filled triangles those by \citet{haering_rix04}. 
The dashed-dotted lines corresponds to a 1.12 slope (the best fit to the data by \citealp{haering_rix04}) 
and the dashed lines show a scatter of $\pm 0.3\,$dex around this relation. 
These lines are the same in all four panels.
They refer to data at $z=0$ but they have also been shown in the diagrams at $z>0$ to ease the comparison.
The results of the AGNICS model are shown for four
different redshifts. 
The formal resolution of the simulation is $2\times 10^{10}M_\odot$ for the galaxy and $4\times 10^{7}M_\odot$ for the black hole.}
\end{figure*}
\begin{figure*}
\noindent
\begin{minipage}{8.6cm}
  \centerline{\hbox{
      \psfig{figure=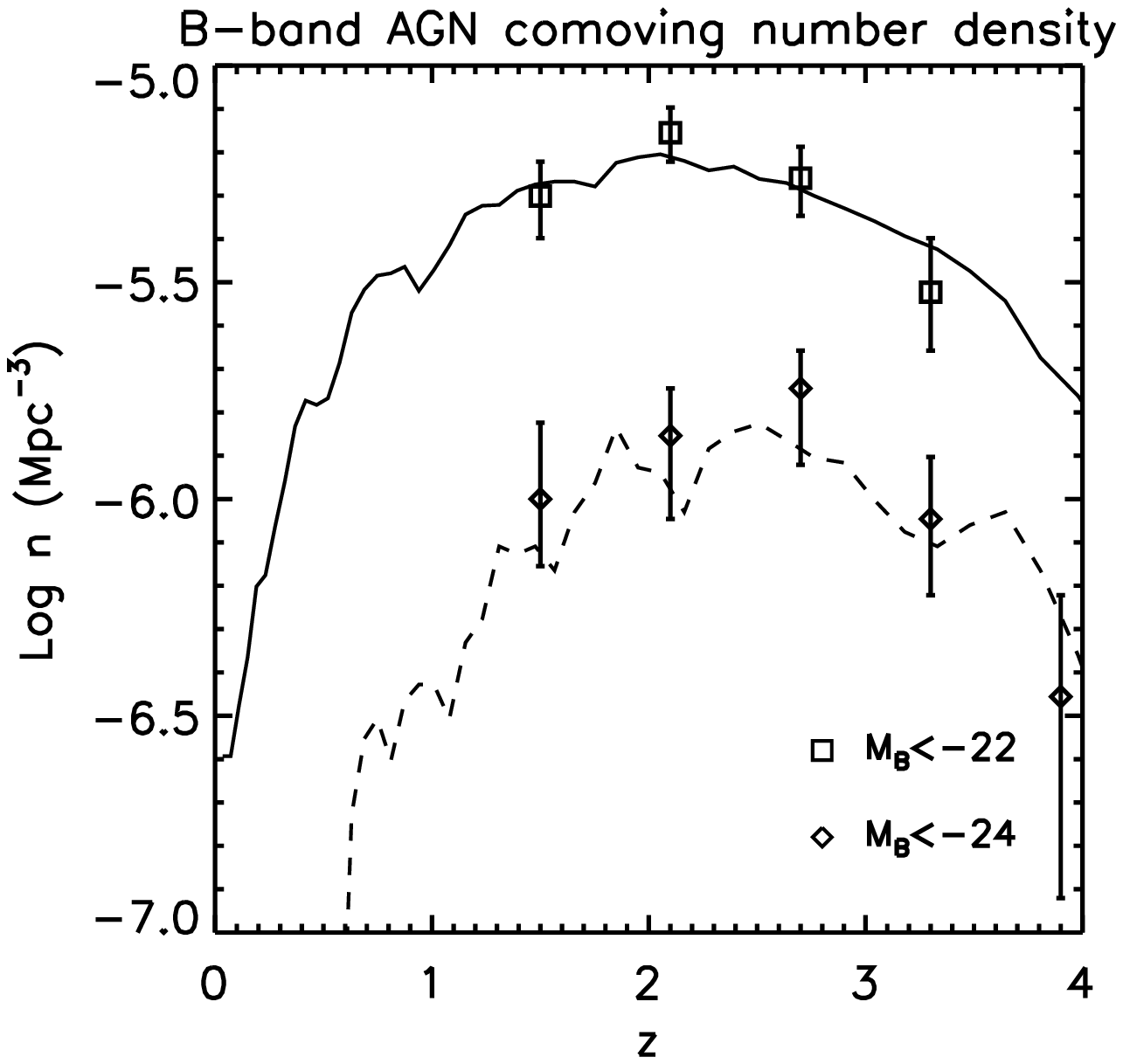,height=8.6cm,angle=0}
  }}
\end{minipage}\    \
\begin{minipage}{8.6cm}
  \centerline{\hbox{
      \psfig{figure=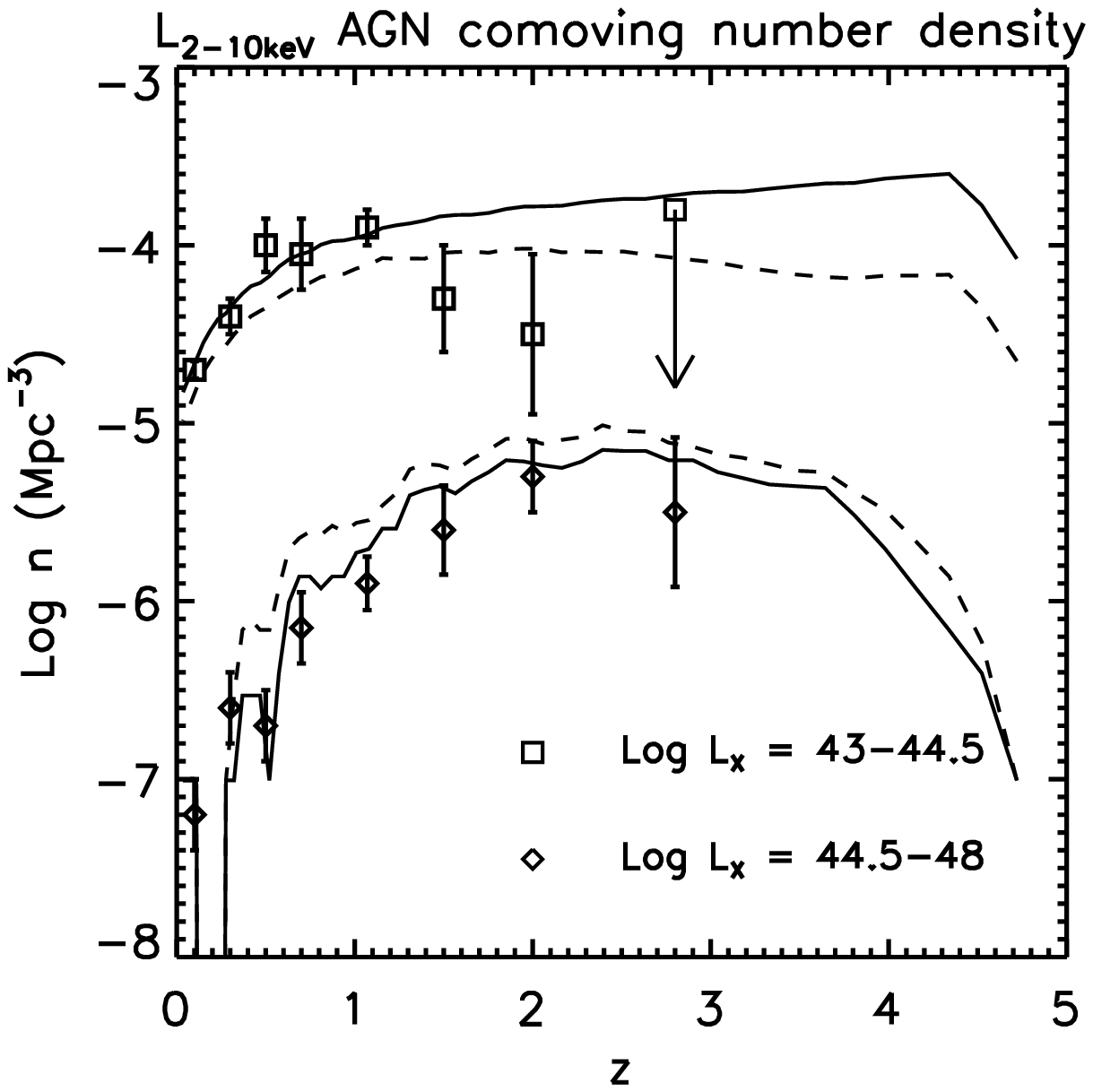,height=8.6cm,angle=0}
  }}
\end{minipage}\    \
\caption{\small Cosmic evolution of the comoving number density of AGN at optical and X-ray wavelengths.
The points with error bars are the data. The optical data are from \citet{wolf_etal03}.
The X-ray data are from \citet{ueda_etal03}. The lines are the results of our reference model.
In the X-ray plot, we have two types of lines: the dashed lines corresponds to a bolometric correction of 32 as in the 
\citet{elvis_etal94} median SED, the solid lines are computed with the luminosity-dependent
bolometric correction in \citet{marconi_etal04}.}
\end{figure*}

The basic model cannot reproduce the observational scatter in any of the two versions. That is not surprising,
because we expect that our approach underestimates the scatter, since it ignores many parameters that can affect the mass accreted
by black holes (Section 3.1).
The real problem comes from the quasar luminosity function (Fig.~5).

Once we have specified the black hole growth model, our only freedom to fit the quasar luminosity function is in
the value of the radiative efficiency parameter $\epsilon_{\rm rad}$ and in 
the choice of the extinction law (the fraction of type 2 AGN as a function of the X-ray luminosity, $f_{\rm obs}(L_X)$).
The luminosity functions corresponding to models A and B in Fig.~5 were computed 
for a radiative efficiency of $\epsilon_{\rm rad}\simeq 0.26$ by using the extinction law II in Fig.~3.
Very likely, these assumptions overestimate the blue light that comes out of powerful AGN.
Nevertheless, Fig.~5 shows that model A underpredicts the number of bright quasars at $z\sim 2$ (model B is even worse).

We can improve the fit to the luminosity function at $z\sim 2$ by pushing the radiative efficiency to an even higher value, 
by removing optical extinction completely and by increasing the normalisation of the $M_\bullet$ - $M_{\rm bulge}$
relation (the data allow an increase up to a maximum of $30\%$).
However, that misses the physical point, besides the difficulty of justifying such assumptions.
The problem exists because the basic model underestimates the scatter in the black hole mass distribution.
If the black hole mass distribution is a Gaussian in Log($M_\bullet$), 
then more scatter in Log($M_\bullet$) with the same mean value of Log($M_\bullet$)
increases the comoving mass density of supermassive black holes, which automatically raises
the total light output of the quasar population.
Therefore, we can fix the lack of bright quasars at $z\simeq 2$ just by introducing a scatter factor
$\propto 10^\chi$ in front of the right hand side of Eq.~(\ref{bar}), where $\chi$ is a Gaussian random deviate.

However, that does not solve the second problem of model A: the simulated evolution of the quasar luminosity function
between $z\sim 2$ and $z\sim 0.5$ is not so strong in the model as it is in the data (here we have made the comparison
with the 2dF data by \citealp{croom_etal04}).
This finding is not surprising because the figures in \citet{kauffmann_haehnelt00} and \citet{cattaneo01} had already shown 
this limitation of the simplest scenario. 
The Durham group too has encountered the same problem when they have started trying and incorporating
AGN into their semi-analytic model of
galaxy formation (R. Malbon, private communication).

\subsection{The reference model}

The method to obtain a model that by construction reproduces the strong evolution of the quasar population
is to identify the parameters that are significantly different in high and low redshift AGN and to force
a dependence of the black hole accretion rate on these parameters.
We find that the main difference between $z\simeq 2$ and $z\simeq 0.5$ AGN is not
in the gas fraction or in the potential well, but in the density of the gas in the central starburst that fuel the AGN.
Therefore, we choose to explore a model in which the black hole accretion rate is
\begin{equation} 
\dot{M}_\bullet\propto\rho_{\rm burst}^\zeta\dot{M}_{\rm *burst}.
\end{equation}
Cattaneo, Haehnelt \& Rees (1999) used a similar approach when they assumed that
$\Delta M_\bullet\propto (1+z)^\eta\Delta M_{\rm *burst}$ with $\eta=2$ in order to obtain a reasonable agreement
with the cosmic evolution of the comoving mass density of supermassive black holes.
The difference is that here $z$ does not appear explicitly. 
We find a reasonable agreement with the local black hole masses and with the luminosity function of quasars
for a dependence with a power of $\zeta=0.5$, a radiative efficiency of
$\epsilon_{\rm rad}=0.1$ and model II for the fraction of obscured AGN (Figs.~5-6, model C, called the reference model hereafter).

\subsubsection{Black hole masses}
The blue (late-type galaxies) and red (early-type galaxies) point clouds in Fig.~6 show the results of
the reference model at different redshifts while the squares and the triangles with error bars are the mass
estimates  by \citet{marconi_hunt03} and \citet{haering_rix04}, and are the same in all four panels.
In fact, they are the same as in Fig.~4.
The black hole mass distribution at $z=2$ is bimodal because two processes fuel AGN:
mergers (predominant in elliptical galaxies) and bar instabilities (predominant in spiral and lenticular galaxies).
Low redshift starbursts are less dense and form objects with lower values of $M_\bullet/M_{\rm bulge}$.
This mechanism generates scatter in the $M_\bullet$ - $M_{\rm bulge}$ relation.
At low redshift the scatter becomes so large that it erases any trace of the original
bimodality and creates a continuity between the products of bars and those of mergers.

The $\rho_{\rm burst}^{0.5}$ factor not only introduces scatter, but also increases the slope of the
Log$(M_\bullet)$ - Log$(M_{\rm bulge})$ relation, since the densest starburst are also the most massive ones.
Perhaps it is not coincidental the data contain a tilt in the same sense (the lines in Fig.~6 show the
$M_\bullet\propto M_{\rm bulge}^{1.12}$ relation found by \citealp{haering_rix04}). 

Since \citet{ferrarese_merritt00} and \citet{gebhardt_etal00}, there has been considerable interest in the $M_\bullet-\sigma$
relation, where $\sigma$ is the velocity dispersion of the host bulge.
This interest has been due to the discovery that the $M_\bullet-\sigma$ relation contains less scatter than the $M_\bullet-L_{\rm bul}$
relation, where $L_{\rm bul}$ is the optical luminosity of the host bulge.
Nevertheless, \citet{marconi_hunt03} and \citet{haering_rix04} have shown that the $M_\bullet-M_{\rm bul}$ is as tight as
the $M_\bullet-\sigma$ relation.
Therefore, there is no reason to believe that the latter is more fundamental than the former.
However, we have chosen to show our simulated $M_{\rm bulge}-\sigma$ relation (Fig.~8).
This appears to have a shallower slope than that inferred from the mass estimates of \citet{ferrarese_merritt00}
and \citet{tremaine_etal02}, but the difference is entirely attributable to the slope in the $M_{\rm bulge}-\sigma$ relation,
which is shallower in GalICS than in the data (see the dashed line in Fig.~8). 
The result is not so bad if we consider that GalICS calculate 
radii, and therefore velocity dispersions, of bulges assuming that mergers conserve the mass and the total energy, while neglecting
the loss of mass and energy in tidal tails and ignoring any angular momentum consideration.

\subsubsection{Evolution of the AGN population}

The fiducial model assumes that $\dot{M}_\bullet\propto\rho_{\rm burst}^{0.5}\dot{M}_{\rm *burst}$.
The scatter introduced by the $\rho_{\rm burst}^{0.5}$ factor is large enough that it allows to fit the luminosity function of quasars
with $\epsilon_{\rm rad}=0.1$ (model C of Fig.~5)
without violating the constraints on black hole masses (Fig.~6).
In Fig.~7, we take exactly the same model and compare its predictions with the data from COMBO-17 \citep{wolf_etal03}, which
probe fainter AGN and higher redshifts (left panel). In the right panel, we show the predictions of our model 
for the comoving number density of AGN with
$10^{43}{\rm\,erg\,s}^{-1}<L_{\rm 2-10\,keV}<3\times 10^{44}{\rm\,erg\,s}^{-1}$ 
and $3\times 10^{44}{\rm\,erg\,s}^{-1}<L_{\rm 2-10\,keV}<10^{48}{\rm\,erg\,s}^{-1}$.
The solid lines and the dashed lines are the predictions derived with the bolometric corrections in \citet{marconi_etal04} and
\citet{elvis_etal94}, respectively.
We compare these predictions with the data by \citet{ueda_etal03} (points with error bars).
The model that uses the bolometric corrections proposed by \citet{marconi_etal04} is in better agreement with the data. 

The fraction of obscured AGN in model C is computed with extinction law II, given by the dashed line in Fig.~3.
We should clearly say that this law has been deliberately constructed to obtain the best possible agreement between
model C and the \citet{croom_etal04} data.
A posteriori, the \citet{barger_etal05} data have shown that this model is not inconsistent with our present knowledge of the
type 2 fraction as a function of luminosity, since
the data points from \citet{ueda_etal03} and \citet{szokoly_etal04} in Fig.~3 do not include the contribution of Compton-thick
AGN to the type 2 fraction (model D shows an alternative version of the reference model, in which we only include the obscuration
that is known to exist from AGN that are detected in X-rays, but do not have UV excess or broad line emission).
One can be worried because the magnitude intervals on which the model and the data are compared at $z\sim 2$ and $z\sim 1$
only overlap at one point ($M_B=-24.25$). Reassuringly, at that point the agreement is good at both redshifts,
but the comparison with the COMBO-17 data in
Fig.~7 gives the best evidence that the success of our model is not a mere consequence of the change in the extinction law at the
magnitude separating high and low redshift data.
Fig.~7 also shows that our model can reproduce a good agreement with the X-ray data.
This agreement is independent of the obscuration model and supports the conclusion that the discrepancy  
between the blue luminosity function without obscuration and the optical data is due to extinction.

We introduced a steep increase in the type 2 fraction at low luminosities to suppress the number of AGN with $M_B>-24$.
\citet{kauffmann_haehnelt00} and \citet{enoki_etal03} dealt with the same problem by introducing a low circular velocity cut-off
(Eqs.~\ref{kh}-\ref{enoki}).
Model F shows our results when we apply this solution in combination with the extinction law I 
derived from the data in \citet{ueda_etal03}.
The fit to the data is good at $z\simeq 2$ and $z\simeq 0.5$. The discrepancy at $z\simeq 1$ occurs because all our AGN
$z\simeq 1$ are in large galaxy groups, where the low circular velocity cut-off has limited consequences,
but this may be an effect of small number statistics due to cosmic variance together with
the finite size of the computational box.
The model with the low circular velocity cut-off gives a steeper slope in the 
Log($M_\bullet$)-Log($M_{\rm bulge}$) relation than the model without this cut-off.

Model E is an alternative version of model C in which we have dropped the $L_{\rm bol}\le L_{\rm Edd}$ condition.
The excess of blue light from super-Eddington AGN is compensated by reducing the radiative efficiency from 
$\epsilon_{\rm rad}=0.1$ to $\epsilon_{\rm rad}=0.08$.
Model E fits the data even better that model C does, but
we have chosen to present model C as our reference model because its assumptions are more standard.

\begin{figure}
\label{bhvsigma} 
\centerline{\psfig{figure=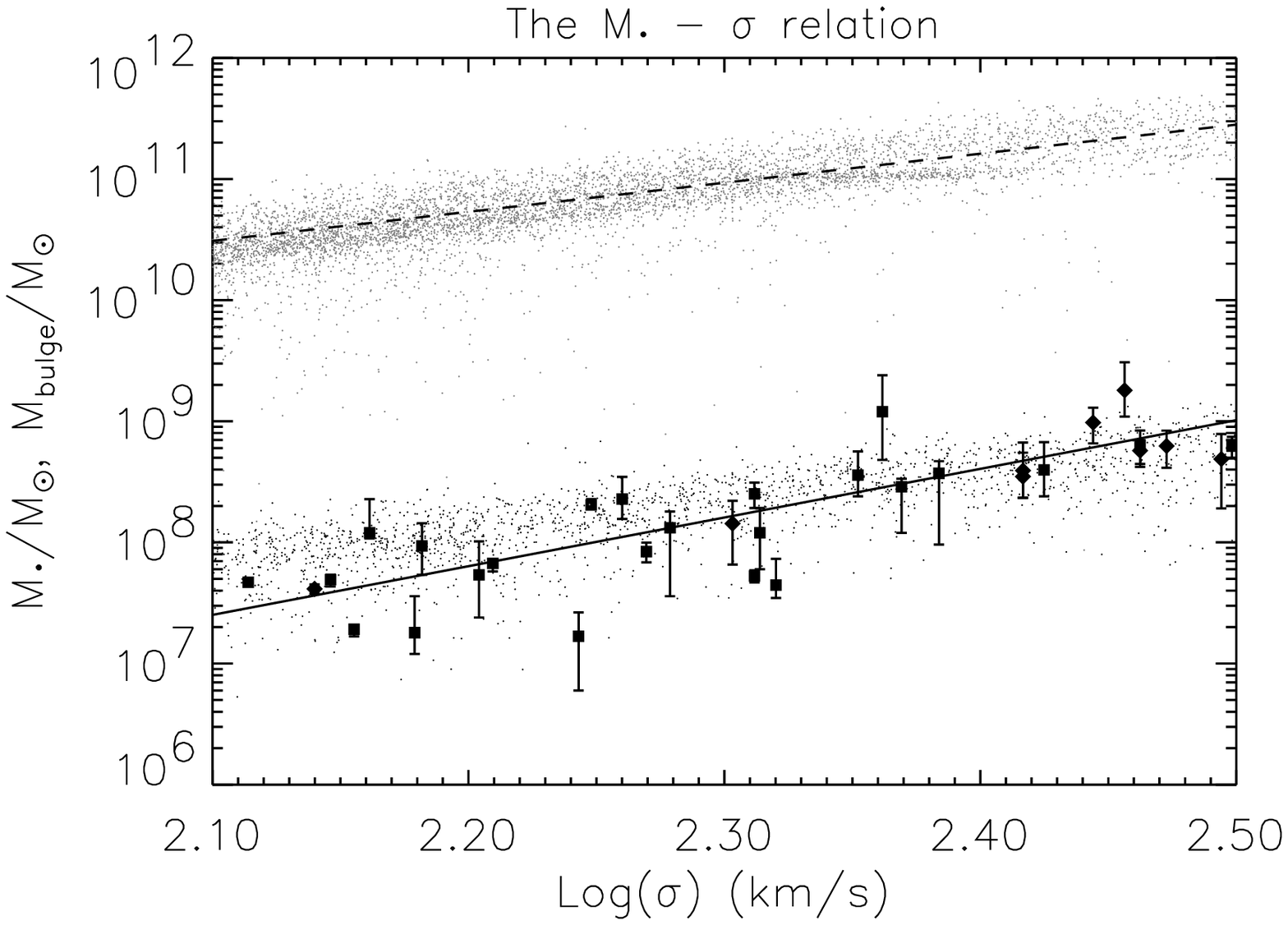,width=0.48\textwidth,angle=0}}
\caption{\small Black hole and bulge mass 
plotted against the velocity dispersion of the host bulge at the half mass radius.
The squares with error bars are the black hole mass estimates of \citet{tremaine_etal02} while the diamonds
with error bars are those of \citet{ferrarese_merritt00}. The continuous solid line is a fit to the data of 
\citet{tremaine_etal02} and has a slope of 4.02. The dashed line has a slope of 2.4. It is not a fit and is only shown to
guide the eye.}
\end{figure}
\begin{figure*}
\noindent
\begin{minipage}{8.6cm}
  \centerline{\hbox{
      \psfig{figure=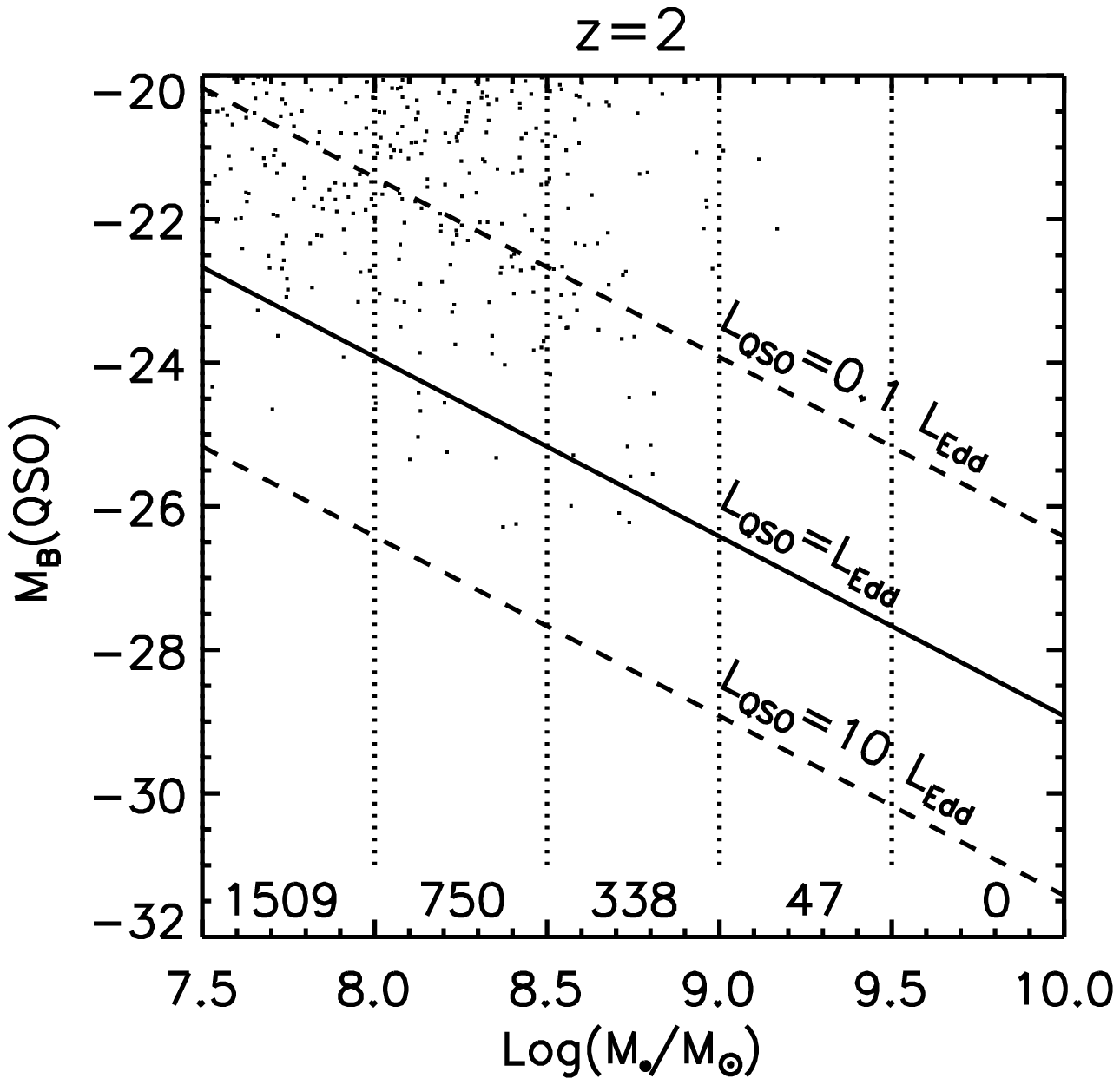,height=8.6cm,angle=0}
  }}
\end{minipage}\    \
\begin{minipage}{8.6cm}
  \centerline{\hbox{
      \psfig{figure=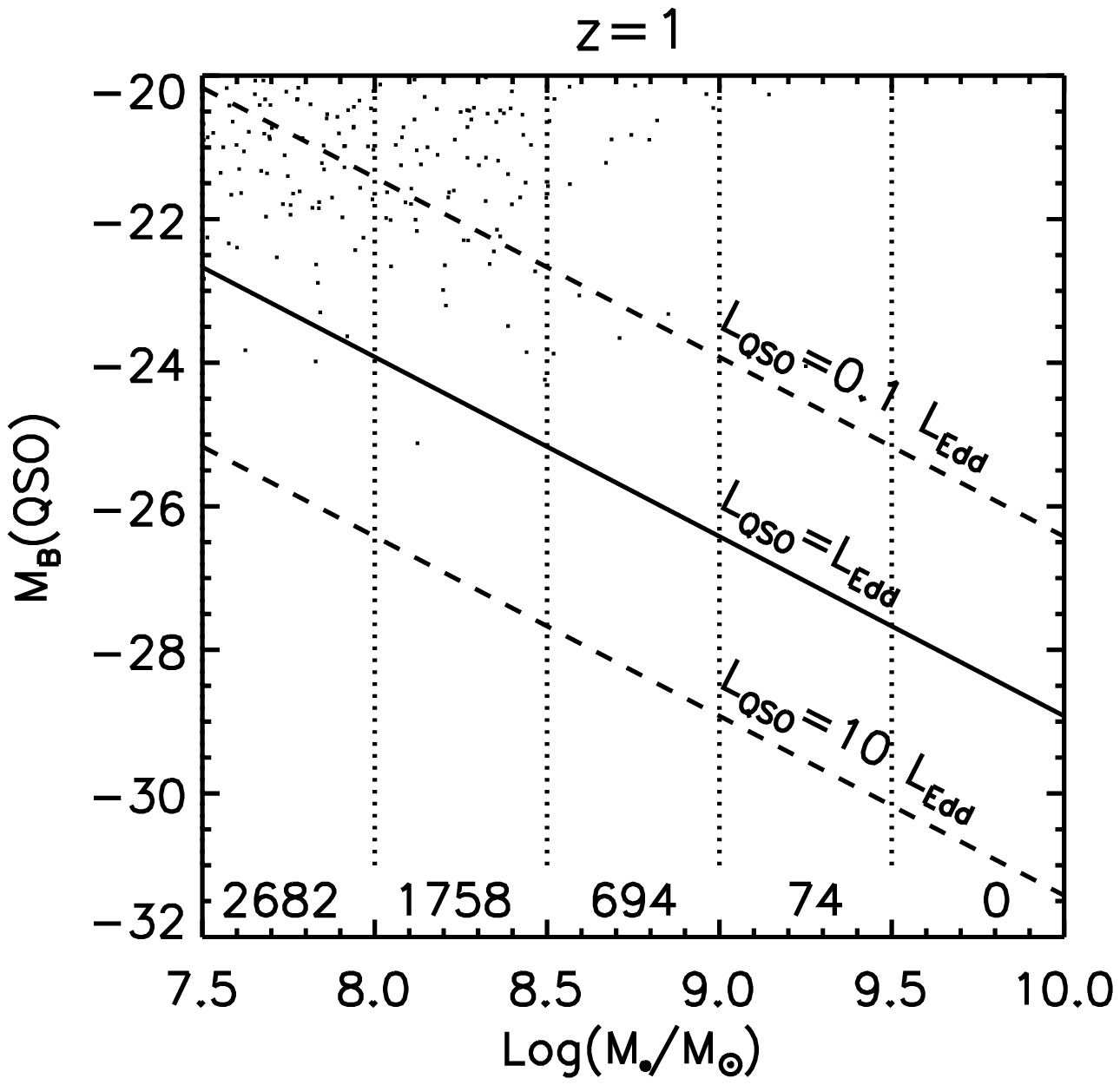,height=8.6cm,angle=0}
  }}
\end{minipage}\    \
\begin{minipage}{8.6cm}
  \centerline{\hbox{
      \psfig{figure=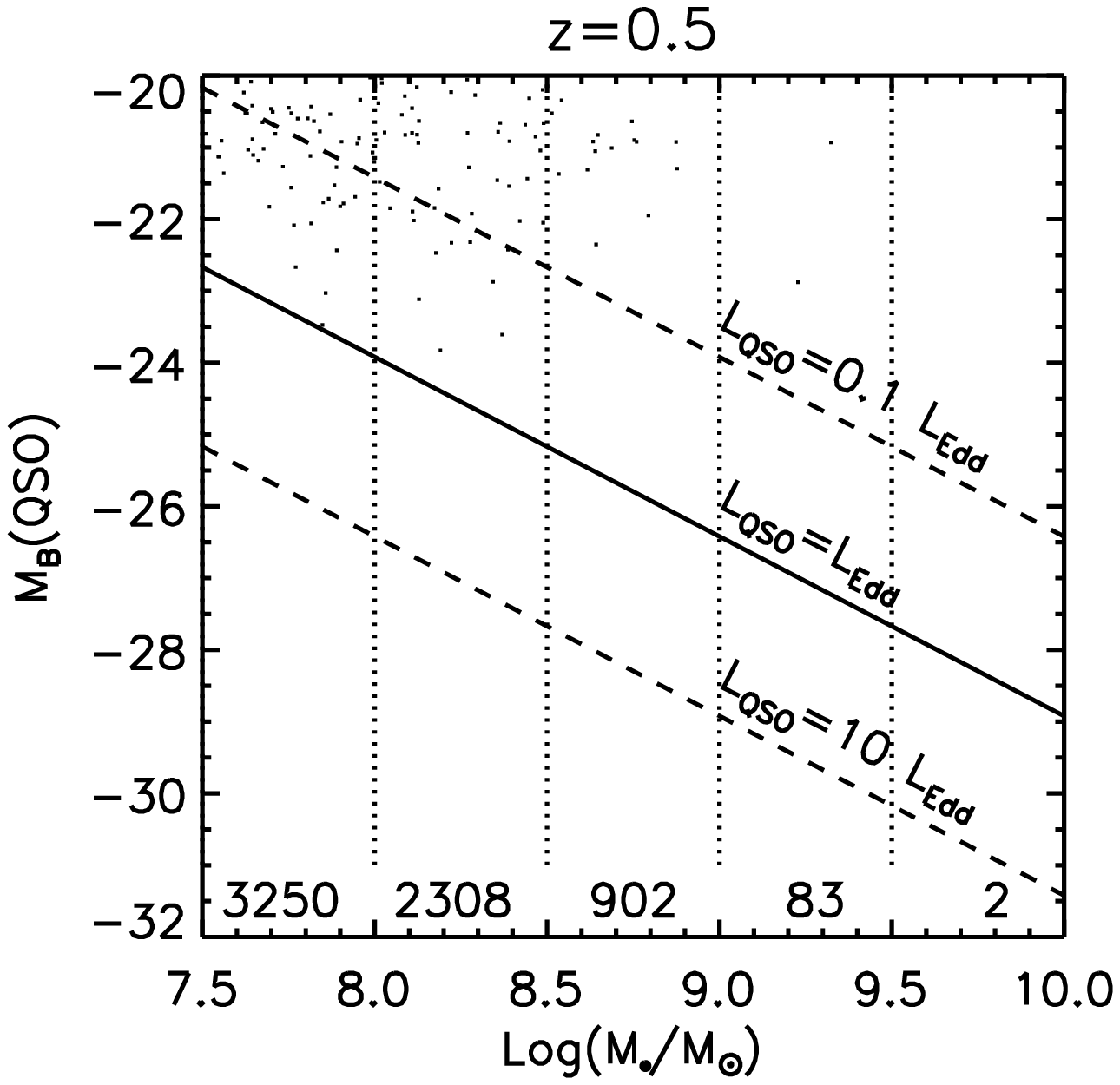,height=8.6cm,angle=0}
  }}
\end{minipage}\    \
\begin{minipage}{8.6cm}
  \centerline{\hbox{
      \psfig{figure=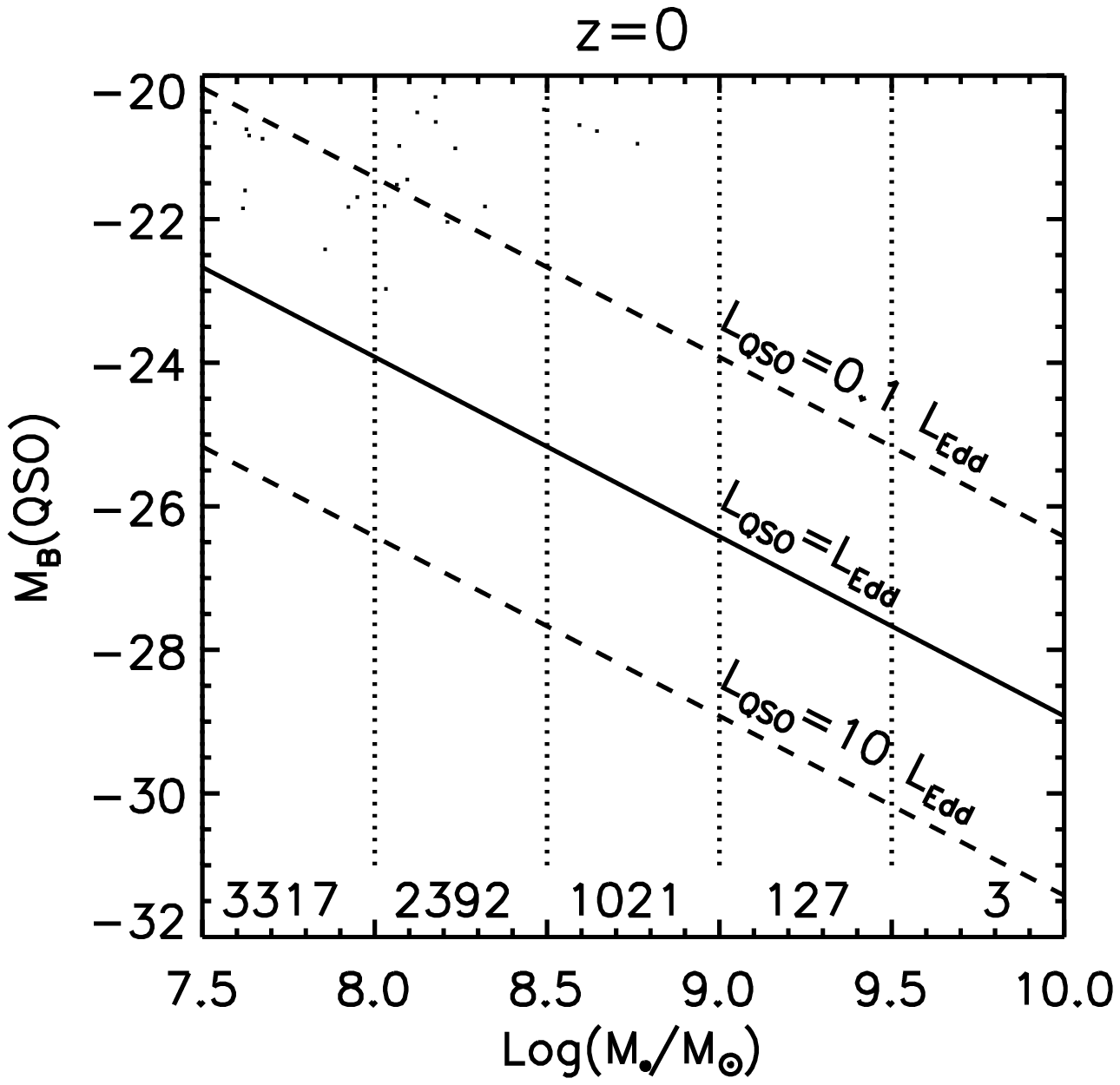,height=8.6cm,angle=0}
  }}
\end{minipage}\    \
\caption{\small The black hole mass - blue luminosity relation in model E
(the reference model without the Eddington limit) at four different redshifts.
Each diagram is divided in six intervals of black hole mass. The row of numbers in the lower part of each diagram gives the 
number of black holes in each mass interval for the (150\,Mpc)$^3$ box with 256$^3$ particles in the dark matter simulation.
The formal resolution for the black hole mass is $\sim 4\times 10^7M_\odot$.}
\end{figure*}
\begin{figure*}
\noindent
\begin{minipage}{8.6cm}
  \centerline{\hbox{
      \psfig{figure=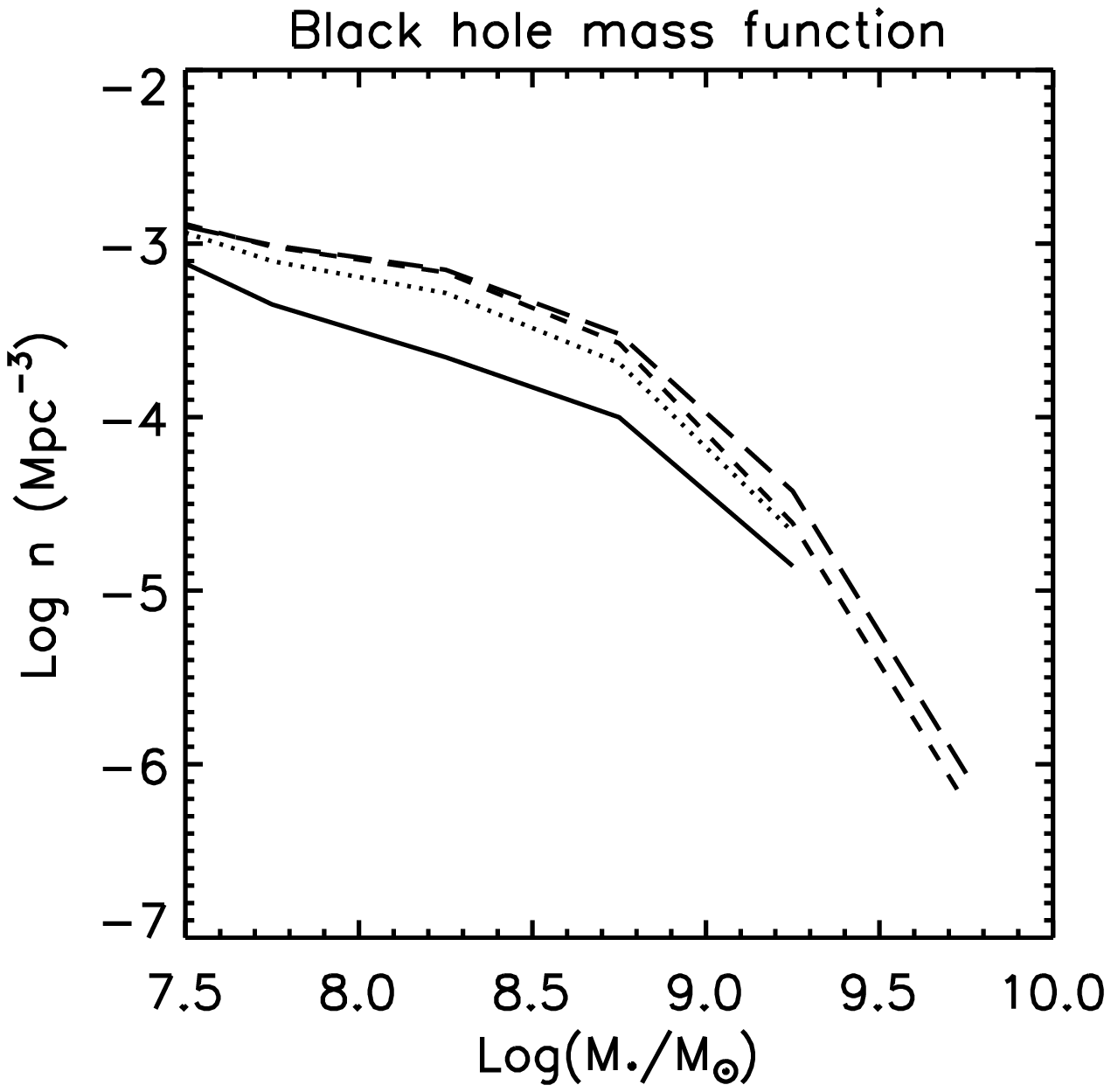,height=8.6cm,angle=0}
  }}
\end{minipage}\    \
\begin{minipage}{8.6cm}
  \centerline{\hbox{
      \psfig{figure=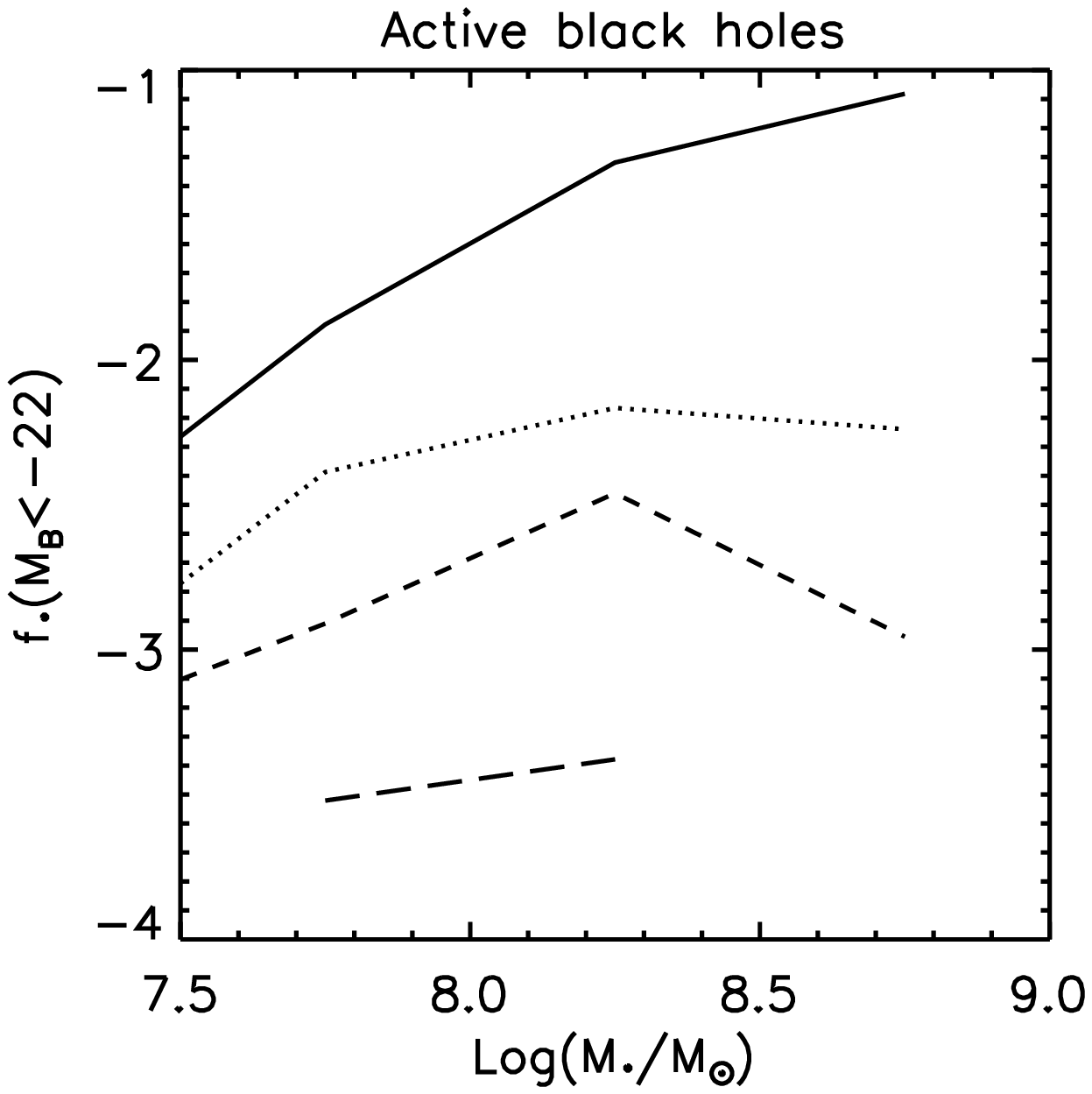,height=8.6cm,angle=0}
  }}
\end{minipage}\    \
\caption{\small Left panel :
cosmic evolution of the black hole mass function in the reference model.  
Right panel: the fraction of black holes with $M_B<-22$ as a function of black hole mass and redshift.
In both panels, the solid, dotted, short-dashed and long-dashed lines 
correspond to $z=$2, 1, 0.5, 0, respectively.
The formal resolution for the black hole mass is $\sim 4\times 10^7M_\odot$.}
\end{figure*}
\subsubsection{AGN activity and black hole mass}

Fig.~9 shows relation between the mass of the black hole and the luminosity of the AGN in model E (the reference model
without the $L_{\rm bol}\le L_{\rm Edd}$ condition).
This figure prompts two considerations.
The first one is that, even when we remove the Eddington limit, it is very difficult to find quasars brighter than
$3-4\,L_{\rm Edd}$.
The second one is the limited statistics deriving from the size of the computational box.
At $z=2$, there are 22 quasars with $M_B<-24$ and only 4 with $M_B<-26$.

Fig.~10 presents the cosmic evolution of the mass function of supermassive black holes and the fraction of black holes
that are active at an $M_B<-22$ level as a function of black hole mass.
These plots are identical for models C and E.
The black hole mass function shows that black holes are still growing at $1<z<2$, but the growth is small at $z<1$.
The evolution in the interval $1<z<2$ is stronger at $M_\bullet\sim 3\times 10^8M_\odot$ than it is at $M_\bullet>10^9M_\odot$.
The fraction of active black holes decreases by almost two orders of magnitude 
between $z\simeq 2$ and $z\simeq 0$.
The most active black holes are objects of $\sim 6\times 10^8M_\odot$ at $z\simeq 2$ 
and $\sim 2\times 10^8M_\odot$ at $z\simeq 0.5$.
The two panels combined suggest a picture in which the most massive black holes form earlier
while less massive black holes continue growing to lower redshifts.
This is consistent with the finding that the most massive galaxies form at higher redshifts 
\citep{monaco_etal00,granato_etal01,corbin_vacca02,cattaneo_bernardi03} also known as down-sizing
of galaxy formation or anti-hierarchical evolution of the baryons with respect to the dark matter,
although Figs.~1 and 10 show that hierarchical models of galaxy formation can in fact reproduce this behaviour.

\section{Discussion and conclusion}

Observational and theoretical arguments (see the Introduction) suggest that the same physical mechanisms are
responsible for the growth of bulges and supermassive black holes.  The main questions are what are these mechanisms and if 
they work one-way only (from the galaxy to the black hole) or both ways (through AGN feedback).

Mergers and disc instabilities provide a path for transforming late-type galaxies into early-type galaxies
while driving a sudden fuel supply into the galactic nucleus. 
The importance of this process is demonstrated by hydrodynamic simulations
and observed in the real Universe.
Semi-analytic models of galaxy formation have incorporated the merger model since the early 1990s and have achieved 
a substantial degree of success (e.g. Fig.~1), but some discrepancies 
(overcooling at the centre of massive haloes, late formation of massive galaxies, low SCUBA counts in relation
to the blue light that comes out of local galaxies) remain.
Critics of semi-analytic models have argued that the great complexity of the 
models and the inherent large number of free parameters allow for the possibility that incorrect assumptions
may be reconciled with the data.
After all, the Ptolemaic model allowed to compute the ephemerides with reasonable accuracy, 
but its basic assumptions were false.
This is a real danger when models are developed to reproduce a small number of observational constraints.
However, the increasing volume of astronomical data from sub-millimetre, infrared, optical and X-ray bands
heavily outnumbers the free parameters in the hand of the simulators. 
The presence of the same discrepancies in models developed independently and the difficulty that the
modellers are encountering in solving these problems prove that the semi-analytic approach is robust
(although some relevant physics are still missing).

In this paper, we have presented AGNICS (Active Galactic Nuclei In Cosmological Simulations), a hybrid approach that incorporates large
cosmological N-body simulations, a semi-analytic model of galaxy formation and a scheme for the growth of supermassive black holes.
We have used this model to investigate a scenario where mergers and disc instabilities
drive massive gas inflows from galaxy discs into compact central starbursts, while
a small fraction of this gas fuels the growth of a supermassive black hole.
If $\dot{M}_\bullet$ and $\dot{M}_{\rm *burst}$ are simply related by a constant of proportionality,
that leaves us with a tight $M_\bullet\propto M_{\rm bulge}^q$ relation with $q\lsim 1$.
In a standard semi-analytic model, this simple scenario is not consistent with the strong evolution of the 
quasar population at $z\lsim 2$ observed in optical studies of the quasar luminosity function (e.g. \citealp{croom_etal04} and
references therein).

Recently, \citet{barger_etal05} have plotted the evolution with redshift of the rest-frame 2-8\,keV comoving energy density production
rate for AGN with $L_{\rm 2-8\,kev}>10^{42}{\rm erg\,s}^{-1}$ and shown that this has a clear peak at $z\sim 1$.
This result goes in the same sense of an earlier study by \citet{cattaneo_bernardi03}, where we assumed that supermassive black holes 
form at the same time as the stellar populations of 
their host galaxies and used this hypothesis to calculate the evolution with redshift of the 
total mass accreted by supermassive black holes per comoving volume and unit time. We found a peak at $z\sim 1.6$, but this can be 
an overestimate if some stars have formed before the black hole.

The discrepancy between X-ray and optical data is attributed to the presence of an obscured AGN population (type 2 AGN),
which predominates at low luminosities \citealp{barger_etal05,szokoly_etal04,ueda_etal03}.
Low redshift AGN are less powerful and therefore more heavily obscured, with the consequence of amplifying the perceived evolution
at optical wavelengths.

Following the method of \citet{soltan82} and \citet{chokshi_turner92}, \citet{yu_tremaine02} 
estimated the mass per unit cosmic volume accreted by optical quasars over the life of the Universe, $\rho_{\bullet B}$,
by using the equation
\begin{equation}
\label{soltan}
\rho_{\bullet B}=C_B{1-\epsilon_{\rm rad}\over\epsilon_{\rm rad}{\rm c}^2}\int{\rm d}t\int L_B\phi(L_B,t){\rm d}L_B,
\end{equation}
where $\phi(L_B,t){\rm d}L_B$ is the number density of quasars with blue luminosity between $L_B$ and $L_B+{\rm d}L_B$
at a time $t$ after the Big Bang and $C_B\equiv L_{\rm bol}/L_B$ (this equation uses Eq.~\ref{bol} in the opposite direction).
For $\epsilon_{\rm rad}=0.1$, $C_B=11.8$ (as in \citealp{elvis_etal94})
and the \citet{boyle_etal00} model for the luminosity function of quasars, 
in which $\phi(L_B,t)=\phi_*(l^a+l^b)$, where $l\equiv L_B/L_B(t)$ and $\phi_*$, $a$ and $b$ are parameters determined from the data,
\citet{yu_tremaine02} found $\rho_{\bullet B}\simeq 2.1\times 10^5M_\odot{\rm Mpc}^{-3}$.
They compared this value to their estimate of the local mass density of supermassive black holes, 
$\rho_\bullet\simeq (2.5\pm 0.4)\times 10^5M_\odot{\rm Mpc}^{-3}$ and concluded that the need for non-luminous accretion is 
limited.
There are two objections to this argument.
The first one is that inserting $\phi(L_B,t)=\phi_*(l^a+l^b)$ into Eq.~(\ref{soltan}) extrapolates the power law behaviour outside the
measured magnitude range. \citet{barger_etal05} argue that this leads to
overestimating $\rho_{\bullet B}$ by a factor of $\sim 1.8$.
Secondly, most other studies \citep{salucci_etal99,merritt_ferrarese01,marconi_etal04} favour values 
closer to $\rho_\bullet\sim 4\times 10^5M_\odot{\rm Mpc}^{-3}$.
It seems plausible to conclude that $50-75\%$ of black hole accretion is optically obscured.
\citet{sazonov_etal04} analysed the combined infra-red and X-ray background and reached a similar conclusion.

With these considerations in mind, we ask ourselves if obscuration, which is more effective on the low redshift luminosity function, 
can reconcile the simple mode 
$\dot{M}_\bullet\propto\dot{M}_{\rm *burst}$ with the data.
The answer is that it cannot because model A in Fig.~5 does not fit the data even if it includes a large amount of obscuration
(which may be an overestimate of the type 2 fraction at low luminosities).
 
There are two possible solutions.
The first one is that this problem is related to the other shortcomings of semi-analytic models of galaxy formation.
Overcooling in massive haloes, late formation of elliptical galaxies, low number of high redshift sub-millimetre sources,
delayed quasar epoch, all point to a scenario in which the most massive baryonic structures form too late.
In recent times, there has been great interest on mechanical (jets, winds) and radiative (Compton heating) 
AGN feedback as a possible answer to this physical problem
\citep{tabor_binney93,binney_tabor95,tucker_david97,ciotti_ostriker97,ciotti_ostriker01,quilis_etal01,reynolds_etal01,reynolds_etal02,basson_alexander03,omma_etal04,omma_binney04}.
In this case, quasars and galaxy formation cease to exist as separate problems because we cannot solve one without solving the other.
We plan to present and discuss such a model in a forthcoming publication of the AGNICS series.

The second possibility, which we have studied in this paper, is to introduce a second parameter that breaks the strict proportionality 
between black hole accretion and star formation.
This solution introduces a scatter and a tilt in the $M_\bullet$ - $M_{\rm bulge}$ relation.
 
A model in which the black hole accretion rate is enhanced 
by a factor proportional to the square root of the gas density can reproduce the evolution of the quasar population over the entire 
redshift range covered by 2dF \citep{croom_etal04} and Combo-17 \citep{wolf_etal03}
data both at optical and X-ray \citep{ueda_etal03} wavelengths, without need to deviate from the
standard radiative efficiency of $\epsilon_{\rm rad}=0.1$ and in a manner that is consistent
with the observational constraints on the normalisation, the slope and the scatter of the 
$M_\bullet$ - $M_{\rm bulge}$ relation.
This model produces a final black hole mass density of $\rho_\bullet\simeq 4.4\times 10^5M_\odot{\rm Mpc}^{-3}$, very close
to the most recent estimate of $4.5\times 10^5M_\odot{\rm Mpc}^{-3}$ by \citet{marconi_etal04} (see the discussion above).
Black holes with $M_\bullet\gsim 1-2\times 10^8M_\odot$ are in bulge dominated galaxies and have mostly grown through mergers.
At $M_\bullet<10^8M_\odot$ disc instabilities are an important growth path.
The good agreement with the X-ray data support the conclusion that the high type 2 fraction that we need to invoke to fit the optical
data at low luminosities has a physical basis.
Allowing or preventing radiation at $L_{\rm bol}>L_{\rm Edd}$ does not any significant difference, which cannot be
compensated by a small change of the value of the radiative efficiency  $\epsilon_{\rm rad}$.
\begin{figure}
\label{AGNpower} 
\centerline{\psfig{figure=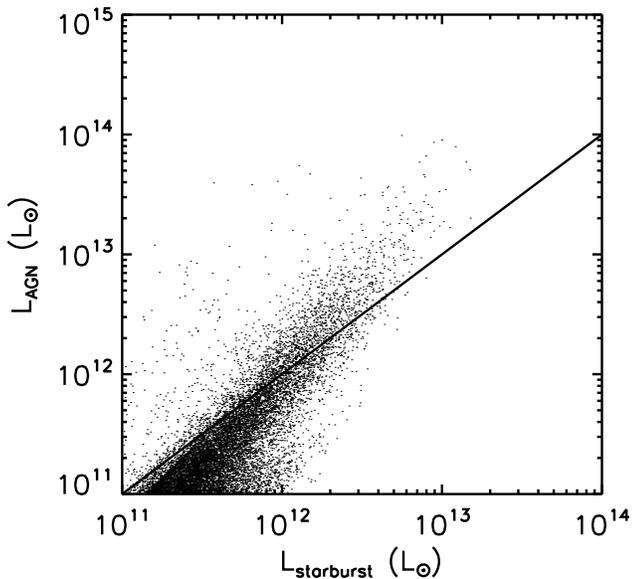,width=0.48\textwidth,angle=0}}
\caption{\small The point cloud shows the relation between the bolometric luminosity of the AGN and the bolometric luminosity of
the starburst component due to star formation only. This relation has been plotted using all the AGN present in the 
reference simulation at all timesteps. The solid line corresponds to $L_{\rm AGN}=L_{\rm starburst}$.}
\end{figure}
It is difficult to say to which extent this model is physical and to which extent the factor $\propto\rho_{\rm burst}^{0.5}$
simply compensates the shortcomings mentioned above. 
Up to a point, it certainly does, but it would be a coincidence if the dependence that compensates a shortcoming of the galaxy formation
model and gives the appropriate evolution of the quasar luminosity function was also the dependence that introduces the right scatter
in the black hole mass distribution (and tilts the distribution of an amount that does not exceed the constraints on the power of
the $M_\bullet$ - $M_{\rm bulge}$ relation). 

The solution proposed in this paper, which breaks the exact proportionality between black accretion and star formation
in favour a higher accretion rate in dense high redshift starbursts, makes two observable predictions.
The first one is that the quasar contribution to the total bolometric luminosity is stronger in the most powerful, denser starbursts
(Fig.~11).
This is a prediction that can be tested by inspecting the SEDs of ultra-luminous infra-red galaxies (ULIRGs)
for AGN features and by studying
how the presence and the strength of these features increase with the total bolometric luminosity of these objects.
\citet{tran_etal01} studied the mid-infrared spectra of 16 ULIRGs and found a transition from mostly
starburst-powered to mostly AGN-powered objects at a luminosity of $\sim 2-3\times 10^{12}\,L_\odot$
in reasonable agreement with the prediction of the reference model (Fig.~11).
The systematic discussion of the infra-red and sub-millimetre properties  of AGN is left to a future publication of the AGNICS series.
We can anticipate that we have checked the most fundamental constraints. 
E.g. even if we make the extreme assumption that
all the AGN power in the Universe is absorbed and reemitted in the far infrared by dust in the host galaxies,
that could account for $<1/3$ of the number of SCUBA sources counted at a flux of $S_{850\mu{\rm m}}>2{\rm\,mJy}$.
The second prediction of our model
is that, for a given bulge mass, the most massive black holes are found in the bulges with the oldest stellar 
populations. 
Observations \citep{merrifield_etal00} seem to confirm this second prediction, too, 
but it is important to have more data to make this conclusion more secure.
Moreover, low redshift mergers have much lower gas fraction than high redshift mergers.
That can be another explanation why bulges with younger stellar populations 
have a lower black hole mass fraction.
The first prediction is thus a cleaner test of our assumption.

\section{Acknowledgements}

A. Cattaneo acknowledges financial support from the European Commission, through a Marie Curie Research Fellowship, and from the Golda Meir Foundation.
He also wishes to thank A. Dekel, G. Mamon and L. Wisotzki for interesting discussion.

\bibliographystyle{mn2e}
\bibliography{references}

\end{document}